\documentclass[final,letterpaper,10pt,journal]{IEEEtran} 

\usepackage{amsmath,amssymb,amsfonts,mathtools}
\usepackage{graphicx}
\usepackage{caption}
\usepackage{subcaption}
\usepackage{cite}
\usepackage{color}
\usepackage{enumerate}
\usepackage{tikz}
\usetikzlibrary{intersections}
\usetikzlibrary {arrows.meta}
\usetikzlibrary {decorations.pathmorphing}
\usetikzlibrary{trees}
\usepackage[justification=centering]{caption}
\usepackage{makecell}
\usepackage[mathscr]{euscript}
\usepackage{stmaryrd}
\usepackage{multirow}
\usepackage{bm,upgreek}
\usepackage{framed}
\usepackage{diagbox}
\usepackage{algorithm,algorithmic}
\usepackage{textcomp,enumitem}
\usepackage{changepage}

\newcommand{\Cov}{\mathrm{Cov}}
\newcommand{\trace}{\mathrm{tr}}
\newcommand{\diag}{\mathrm{diag}}
\newcommand{\blkdiag}{\mathrm{block\textrm{-}diag}}
\newcommand{\rank}{\mathrm{rank}}

\newcommand{\E}{\mathbb{E}}
\newcommand{\diff}{\mathrm{d}}
\newcommand{\Trans}{\mathrm{T}}

\def\MatrixFont{\bf}
\def\VectorFont{\bf}

\newcommand{\mA}{{\MatrixFont A}}
\newcommand{\mB}{{\MatrixFont B}}
\newcommand{\mC}{{\MatrixFont C}}

\newcommand{\mF}{{\MatrixFont F}}
\newcommand{\mG}{{\MatrixFont G}}

\newcommand{\mI}{{\MatrixFont I}}

\newcommand{\mM}{{\MatrixFont M}}

\newcommand{\mP}{{\MatrixFont P}}
\newcommand{\mQ}{{\MatrixFont Q}}
\newcommand{\mR}{{\MatrixFont R}}
\newcommand{\mS}{{\MatrixFont S}}

\newcommand{\va}{{\VectorFont a}}

\newcommand{\vd}{{\VectorFont d}}

\newcommand{\vu}{{\VectorFont u}}
\newcommand{\vv}{{\VectorFont v}}
\newcommand{\vw}{{\VectorFont w}}
\newcommand{\vx}{{\VectorFont x}}
\newcommand{\vy}{{\VectorFont y}}

\newcommand{\vones}{{\VectorFont 1}}

\def\R{\mathbb R}
\def\N{\mathcal N}
\def\dx{n_x}
\def\dy{n_{y_i}}
\def\dd{n_d}
\def\nS{M}
\def\vmu{\bm \mu}
\def\vnu{\bm \nu}
\def\vomega{\bm \omega}
\def\mSigma{\bm \Sigma}

\def\mUpsilon{\bm \Upsilon}
\def\Adj{\rm Adj}
\newcommand{\cen}{\rm{cen}}
\newcommand{\pre}{\rm{update}}
\newcommand{\update}{\rm{(A2)}}
\def\sA{\mathbb A}
\def\sS{\mathbb S}

\newtheorem{theorem}{Theorem}
\newtheorem{lemma}{Lemma}
\newtheorem{proposition}{Proposition}
\newtheorem{remark}{Remark}

\newtheorem{definition}{Definition}
\newtheorem{assumption}{Assumption}

\renewcommand{\IEEEQED}{\IEEEQEDopen}
\IEEEoverridecommandlockouts
\allowdisplaybreaks[3]
\begin{document}
\title{Distributed Fusion Estimation with Protecting Exogenous Inputs
\thanks{
	The work was supported by National Natural Science Foundation of China under Grants 62025306, 62433020, 62203045 and T2293770, CAS Project for Young Scientists in Basic Research under Grant YSBR-008,  and the Talent Scientific Fund of Lanzhou University under Grant 561120225. The material in this paper was not presented at any conference. \emph{(Corresponding author: Ji-Feng Zhang)}}
\thanks{Liping Guo is with the School of Mathematics and Statistics, Lanzhou University, Lanzhou 730000, China. (e-mail: lipguo@outlook.com)}
\thanks{Jimin Wang is with the School of Automation and Electrical Engineering, University of Science and Technology Beijing, Beijing 100083, and also with the Key Laboratory of Knowledge Automation for Industrial Processes, Ministry of Education, Beijing
	100083, China. (e-mail: jimwang@ustb.edu.cn)}
\thanks{Yanlong Zhao is with the Key Laboratory of Systems and Control, Institute of Systems Science, Academy of Mathematics and Systems Science, Chinese Academy of Sciences, Beijing 100190, China. (e-mail: ylzhao@amss.ac.cn)}
\thanks{Ji-Feng Zhang is with the School of Automation and Electrical Engineering, Zhongyuan University of Technology, Zheng Zhou 450007; and also with the Key Laboratory of Systems and Control, Institute of Systems Science, Academy of Mathematics and Systems Science, Chinese Academy of Sciences, Beijing 100190, China. (e-mail: jif@iss.ac.cn)}}
\author{Liping Guo, \IEEEmembership{Member,~IEEE}, Jimin Wang, \IEEEmembership{Member,~IEEE}, Yanlong Zhao,~\IEEEmembership{Senior Member,~IEEE},
and Ji-Feng Zhang,~\IEEEmembership{Fellow,~IEEE}}

\maketitle
\IEEEpeerreviewmaketitle
\thispagestyle{plain}

\begin{abstract}
In the context of distributed fusion estimation, directly transmitting local estimates to the fusion center may cause a privacy leakage concerning exogenous inputs. Thus, it is crucial to protect exogenous inputs against full  eavesdropping while achieving distributed fusion estimation.   
To address this issue, a noise injection strategy is provided by injecting mutually independent noises into the local estimates transmitted to the fusion center. 
To determine the covariance matrices of the injected noises,
a constrained minimization problem is constructed by minimizing the sum of mean square errors of the local estimates while ensuring ($\epsilon, \delta$)-differential privacy.
Suffering from the non-convexity of the minimization problem, an approach of relaxation is proposed, which efficiently solves the minimization problem without sacrificing differential privacy level.
Then, a differentially private distributed fusion estimation algorithm based on the covariance intersection approach is developed. 
Further, by introducing a feedback mechanism, the fusion estimation accuracy is enhanced on the premise of the same ($\epsilon, \delta$)-differential privacy. 
Finally, an illustrative example is provided to demonstrate the effectiveness of the proposed algorithms, and the trade-off between differential privacy level and fusion estimation accuracy.
\end{abstract}

\begin{IEEEkeywords}
	Differential privacy, distributed fusion estimation, constrained optimization, exogenous inputs, full eavesdropping
\end{IEEEkeywords}

\section{Introduction}

The real-time state estimation problem aims at estimating system state from noisy measurements and plays an important role in many areas, such as target tracking and aerospace engineering \cite{Bar-Shalom-Li-KirubarajanEstimation2001,Yan-2023-Distributed,Yan-2023-Guaranteeing}.
In contrast to single-sensor state estimation, multi-sensor fusion estimation employing multi-source data improves accuracy and robustness simultaneously, thereby attracting significant attention in recent years \cite{Yan-2023-Distributed,Yan-2023-Guaranteeing,Cao-2023-Distributed,Yang-2022-State}. 
There are two basic networks for multi-sensor fusion estimation, i.e., centralized and distributed networks. 
Compared with the former, the latter stands out due to better robustness and system feasibility.
However, distributed networks are susceptible to many different types of attacks, such as denial-of-service (DoS) attacks \cite{Tian-2022-Event} and false data injection (FDI) attacks \cite{Li-2018-False}.
To defend against DoS and FDI attacks, many meaningful works about distributed fusion estimation have been proposed (see, e.g.,
\cite{Chen-2019-Distributed,Chen-2018-Secure,Liu-2022-Event-Triggered,Ren-2020-Secure,Zhang-2020-Sequential,Song-2019-Attack,Chen-2019-Fusion}).
In these studies, the attackers should acquire some privacy information by eavesdropping before launching a strategic attack \cite{Xu-2023-Distributed}.
Under this case, it is essential to protect privacy information against eavesdropping at its source. 
Thus, it is of great significance to study privacy preservation problem against eavesdropping in distributed fusion estimation.



To defend against eavesdropping,
some distributed fusion estimation approaches have been developed in \cite{Xu-2023-Distributed,Yan-2023-Distributed,Yan-2023-Guaranteeing}. 
Specifically, encryption-based distributed fusion estimation approaches are presented in \cite{Xu-2023-Distributed,Yan-2023-Distributed}. In \cite{Yan-2023-Guaranteeing}, a differentially private distributed fusion estimation algorithm is provided, where the publicly released estimates defined by fusion estimates averaged over time are protected. 
Attributed to its powerful performance and rigorous mathematical models, differential privacy stands out from its competitors and is studied in a wide range of fields, such as federated learning \cite{Wei-2024-Gradient}, consensus \cite{Wang-2024-Consensus}, optimization \cite{Tie-2022-Differentially,Wang-2024-Tailoring,Han-Differentially-2016}, game theory \cite{Ye-2022-Nash,Wang-2024-Differentially} and control theory \cite{Yu-2020-Design,Yazdani-2023-Differentially}. 
Particularly in state estimation fields, differentially  private filtering
has been firstly discussed in \cite{Ny-2014-Filtering}.  
In addition to measurements and state estimates, exogenous inputs may also contain private information, as demonstrated in applications such as smart grids  \cite{Li-2018-Information-Theoretic} and building automation \cite{Nekouei-2022-Optimal,Weng-2023-Optimal,Guo-2024-Privacy}. 
Therefore, protecting exogenous inputs is critically important, yet it introduces distinct theoretical challenges, including establishing privacy condition, designing optimal noise, and co-optimizing privacy and estimation accuracy.
To our best knowledge, research on protecting exogenous inputs in fusion estimation is still lacking. 

Motivated by the above analysis, in this paper, we study the differentially private distributed fusion estimation to protect the exogenous inputs against full eavesdropping. 
To ensure differential privacy of these exogenous inputs, we have to sacrifice some fusion estimation accuracy due to the noise injection strategy adopted at the sensor side. 
Particularly, we aim at minimizing the sum of mean square errors (MSEs) of local estimates and ensuring ($\epsilon, \delta$)-differential privacy simultaneously, which is the main purpose of this paper.
Unfortunately, there exist some substantial difficulties in achieving this goal:
i) To ensure ($\epsilon, \delta$)-differential privacy, a joint consideration of all the local sensors is necessary; under this case, 
minimizing the sum of MSEs of local estimates is challenging,
especially when the correlation among the measurement noises of local sensors is unknown. 
ii) Computational efficiency is critically important in real-time state estimation, but the minimization problem is non-convex with the optimization variables being matrices, greatly increasing the difficulty of solving it efficiently.
iii) Accurate fusion can foster optimal resource utilization and stability in estimation algorithms; thus, is it possible to enhance fusion estimation accuracy while ensuring ($\epsilon, \delta$)-differential privacy?
These difficulties are properly solved in this paper, and the main contributions are summarized as follows:
\begin{adjustwidth}{0pt}{}
\begin{enumerate}[label=\textbullet]
	\item We achieve distributed fusion estimation while protecting exogenous inputs against full eavesdropping. Unlike common differentially private approaches that often inject simple isotropic or scalar noise (see, e.g., \cite{Wei-2024-Gradient,Wang-2024-Consensus,Ny-2014-Filtering,Yan-2023-Guaranteeing,Wang-2024-Tailoring,Tie-2022-Differentially,Ye-2022-Nash,Wang-2024-Differentially}), we propose an optimized anisotropic noise injection strategy tailored to system uncertainties. This strategy introduces less noise along directions where the state is already uncertain, thereby improving estimation accuracy without compromising privacy. The noise covariance matrix is obtained by solving a constrained minimization problem that minimizes the sum of MSEs of local estimates while ensuring ($\epsilon, \delta$)-differential privacy. Furthermore, we solve this problem efficiently via an SDP relaxation and establish an explicit upper bound on the relaxation gap. The proposed SDP not only preserves the privacy guarantee but also meets real-time computation requirements. 
	\item We develop two differentially private distributed fusion estimation algorithms based on covariance intersection, balancing low-complexity and high-accuracy requirements. 
	For the first algorithm, we provide an analytical characterization of the estimation accuracy loss, which rigorously quantifies the privacy-accuracy trade-off. 
	For the second algorithm, we incorporate a feedback mechanism that is theoretically guaranteed to enhance estimation accuracy without compromising the ($\epsilon, \delta$)-differential privacy, but at the expense of increased computational complexity, thereby establishing a  complexity-accuracy trade-off.
\end{enumerate}
\end{adjustwidth}


\textit{\textbf{Notations.}}
Scalars, vectors and matrices are denoted by lowercase letters, bold lowercase letters, and bold capital letters, respectively. 
Scalar $0$, zero vector, and zero matrix are all denoted by $0$ for simplicity.
All the vectors are column vectors.
The set of all $n$-dimensional real vectors and all $n \times m$ real matrices are denoted by $\R^{n}$ and $\R^{n \times m}$, respectively.  
For a vector $\va$, $\|\va\|$ denotes its Euclidean norm, further, $\|\va\|_{\mA}$ denotes its Euclidean norm weighted with $\mA > 0$, i.e., $\sqrt{\va^{\Trans} \mA \va}$. 
$\diag(\va)$ represents the diagonal matrix with $\va$ on the principal diagonal.
In particular, $\vones$ represents the vector with all entries one. 
For a square matrix $\mA$,
$\mA \geq 0$ (or $\mA > 0$) means that $\mA$ is positive semi-definite (or positive definite).
$\trace(\mA)$ represents the trace of $\mA$.
$\lambda_{\rm{min}}(\mA)$ denotes the minimum eigenvalue of $\mA$.
The $\blkdiag(\mA_0, \mA_1, \dots, \mA_n)$ represents the block diagonal matrix with matrices $\mA_0, \mA_1, \dots, \mA_n$ on the principal diagonal. 
$\mA \otimes \mB$  represents the Kronecker product operation between matrices $\mA$ and $\mB$.  
$\mI_n$ represents the $n \times n$ identity matrix.
$\E[\cdot]$ is the mathematical expectation operator.

\section{Problem formulation}





Consider a distributed multi-sensor network system consisting of $\nS$ local sensors and a fusion center. For Node $i = 1, 2, \dots, \nS$, the following time-varying dynamic system is addressed:
\begin{equation}\label{eq:system}
\begin{aligned}
	\vx_{k + 1} &= \mA_k \vx_k + \mB_k \vd_k + \vw_k, \\
	\vy_{i,k} &= \mC_{i,k} \vx_k + \vv_{i,k},
\end{aligned}
\end{equation}
where $k = 0, 1, 2, \dots$ is time index, $\vx_k \in \R^{\dx}$, $\vd_k \in \R^{\dd}$, and $\vy_{i,k} \in \R^{\dy}$ are the state, the exogenous input, and the $i$-th node's measurement, respectively, $\mA_k \in \R^{\dx \times \dx}$, $\mB_k \in \R^{\dx \times \dd}$, and $\mC_{i,k} \in \R^{\dy \times \dx}$ are known matrices, $\{\vw_k\}$ and $\{\vv_{i,k}\}$ are zero-mean Gaussian white noise sequences with covariance matrices $\mQ_k$ and $\mR_{i,k}$, respectively, and $\vd_k$ is regarded as deterministic but unknown  \cite{kitanidis1987unbiased,darouach1997unbiased}.
The initial state is independent of the noise sequences. 
All system parameters, including $\mA_k$, $\mB_k$, $\mC_k$, $\mQ_k$ and $\mR_{i,k}$, are available to the fusion center.
The correlations between the measurement noises of different sensors are typically unknown due to factors such as physical separation and unsynchronized clocks (see, e.g., \cite{Bar-Shalom-Li-KirubarajanEstimation2001,Julier-1997-A}). 

\begin{assumption}\label{assum:full rank}
	$\rank(\mC_{i,k} \mB_{k-1}) = \rank(\mB_{k-1}) = \dd$, for all $k$.
\end{assumption}

\begin{assumption}\label{assum:detectable}
	$(\mA_k, \mC_{i,k})$ is detectable, for all $k$.
\end{assumption}

\begin{remark}
	Assumptions \ref{assum:full rank} and \ref{assum:detectable} are standard in the literature (see, e.g., \cite{kitanidis1987unbiased,darouach1997unbiased}). 
		Assumption \ref{assum:full rank} ensures $\dx \geq \dd$ and $\dy \geq \dd$, while Assumption \ref{assum:detectable} guarantees a bounded error covariance in Kalman filtering.
\end{remark}

Distributed fusion estimation aims to produce a fusion estimate and its associated error covariance matrix at each time step $k$, through the fusion of all local estimates and their error covariance matrices. 

At the sensor side,  the linear unbiased minimum-variance state estimation proposed in \cite{kitanidis1987unbiased} is adopted, which is derived in the minimum mean square error (MSE) sense and comprises two steps. For Node $i = 1, 2, \dots, \nS$, let $\hat \vx_{i,k|k}$ be the estimate of $\vx_k$ using the measurements $\vy_{i,0}, \vy_{i,1}, \dots, \vy_{i,k}$, and $\mP_{i,k|k}$ be the associated error covariance matrix.

1) Prediction step. The predicted state estimate, denoted by $\hat \vx_{i,k|k-1}$, and its error covariance matrix, denoted by $\mP_{i,k|k-1}$, are calculated as follows:
\begin{align}
	\hat \vx_{i,k|k-1} &= \mA_{k-1} \hat \vx_{i,k-1|k-1}, \label{eq:non-privacy:pre:mean}\\
	\mP_{i,k|k-1} &= \mA_{k-1} \mP_{i,k-1|k-1} \mA_{k-1}^{\Trans} + \mQ_{k-1}.\label{eq:non-privacy:pre:cov}
\end{align}

2) Update step. Once receiving the measurement $\vy_{i,k}$, the linear unbiased minimum-variance state estimate and its error covariance matrix are given as
\begin{align}
	\hat \vx_{i,k|k} &= \hat \vx_{i,k|k-1} + \mG_{i,k} (\vy_{i,k} - \mC_{i,k} \hat \vx_{i,k|k-1}), \label{eq:umv:mean}\\
	\mP_{i,k|k} &= \mP_{i,k|k-1} - \mP_{i,k|k-1}  \mC_{i,k}^{\Trans} \mF_{i,k}^{-1} \mC_{i,k} \mP_{i,k|k-1} \notag\\
	&\quad + (\mB_{k - 1} - \mP_{i,k|k-1} \mC_{i,k}^{\Trans} \mF_{i,k}^{-1} \mC_{i,k} \mB_{k-1})  \notag\\
	&\quad \cdot (\mB_{k-1}^{\Trans} \mC_{i,k}^{\Trans} \mF_{i,k}^{-1} \mC_{i,k} \mB_{k-1})^{-1} \notag\\
	&\quad \cdot (\mB_{k-1} - \mP_{i,k|k-1} \mC_{i,k}^{\Trans} \mF_{i,k}^{-1} \mC_{i,k} \mB_{k-1})^{\Trans}, \label{eq:umv:cov}
\end{align}
where
\begin{align}
	\mG_{i,k} &= \mP_{i,k|k-1} \mC_{i,k}^{\Trans} \mF_{i,k}^{-1} \notag\\
	&\quad + (\mB_{k -1} - \mP_{i,k|k-1} \mC_{i,k}^{\Trans} \mF_{i,k}^{-1} \mC_{i,k} \mB_{k -1})  \notag \\
	&\quad \cdot (\mB_{k-1}^{\Trans} \mC_{i,k}^{\Trans} \mF_{i,k}^{-1} \mC_{i,k} \mB_{k-1})^{-1} \mB_{k-1}^{\Trans} \mC_{i,k}^{\Trans} \mF_{i,k}^{-1}, \notag \\
	\mF_{i,k} &= \mC_{i,k} \mP_{i,k|k-1} \mC_{i,k}^{\Trans} + \mR_{i,k}. \notag
\end{align}

\begin{remark}
	It should be pointed out that \eqref{eq:umv:mean} corresponds to Eqs.~(5) and (19) in \cite{kitanidis1987unbiased}, while \eqref{eq:umv:cov} corresponds to Eq.~(21) in \cite{kitanidis1987unbiased}, and \cite{kitanidis1987unbiased} also provides the detailed derivations. Further, the first term of \eqref{eq:umv:cov}, $\mP_{i,k|k-1} - \mP_{i,k|k-1}\mC_{i,k}^{\Trans} \mF_{i,k}^{-1} \mC_{i,k} \mP_{i,k|k-1}$, is the standard Kalman filter update, while the second is a correction that accounts for the uncertainty induced by $\vd_{k-1}$, thereby ensuring the unbiasedness of the state estimate \eqref{eq:umv:mean}.
\end{remark}

At the fusion center side, after receiving all the local state estimates and error covariance matrices, the fusion estimate, denoted by $\hat \vx_{k|k}^{\text{non-priv}}$, and its error covariance matrix, denoted by $\mP_{k|k}^{\text{non-priv}}$, are derived 
by employing the approach of covariance intersection \cite{Julier-1997-A}:
\begin{align*}
	\big(\mP_{k|k}^{\text{non-priv}}\big)^{-1} \hat \vx_{k|k}^{\text{non-priv}} & =\sum_{i = 1}^{\nS} w_i \mP_{i,k|k}^{-1}  \hat \vx_{i,k|k}, \\
	\big(\mP_{k|k}^{\text{non-priv}}\big)^{-1} & =\sum_{i = 1}^{\nS} w_i \mP_{i,k|k}^{-1}, 
\end{align*}
where the weight vector $\vw = [w_1, w_2, \dots, w_{\nS}]^{\Trans}$ satisfies $w_i \geq 0$ and $\vw^{\Trans} \vones = 1$. 

\begin{remark}
	Our usage of the term ``distributed'' is standard and refers to the well-established ``distributed sensing with centralized fusion" architecture  (see, e.g., \cite{Bar-Shalom-Li-KirubarajanEstimation2001}). Unlike a fully distributed peer-to-peer network without a fusion center, our architecture processes raw measurements locally and independently at individual sensors, which then transmit their local estimates to the fusion center.
\end{remark}

For the system \eqref{eq:system}, there may be an eavesdropper, who is external to the system and trying to infer the private information.
We assume that the eavesdropper has the following capability, which is referred to as the full eavesdropping hereinafter.

\begin{definition}[Full eavesdropping]\label{assum:full eavesdropping}
The full eavesdropping is a form of passive wiretapping that silently monitors communication channels without modification, and can obtain the same information as the fusion center, which can be classified into two categories: 
(a) system parameters, including $\mA_k$, $\mB_k$, $\mC_k$, $\mQ_k$ and $\mR_{i,k}$;
(b) real-time transmitted data.

\end{definition}

\begin{figure}[htbp]
	\centering
	\includegraphics[width=0.48\textwidth]{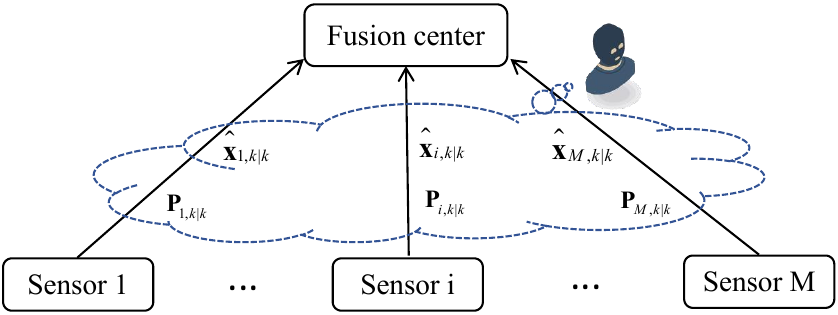}
	\caption{Architecture of the problem: Each sensor $i = 1, 2, \dots, \nS$ completes local estimation task independently, and transmits $\hat \vx_{i,k|k}$ and $\mP_{i,k|k}$ to the fusion center for the fusion estimation task, where all the local estimates $\{\hat \vx_{i,k|k}\}_{i = 1}^{\nS}$ and their error covariance matrices $\{\mP_{i,k|k}\}_{i = 1}^{\nS}$ are available to the full eavesdropper.}
	\label{fig:demo}
\end{figure}

The architecture of the problem is depicted in Fig.~\ref{fig:demo}. 
Note that there is no direct information transmission among the local sensors. 
In the system \eqref{eq:system}, the exogenous input $\vd_{k-1}$ at time step $k$ contains private information and should be protected against full eavesdropping. 

To demonstrate the necessity of protecting $\vd_{k-1}$ at time step $k$, the following two practical examples are representative:
\begin{adjustwidth}{0pt}{}
\begin{enumerate}[label=\textbullet]
	\item \textbf{Smart grids (see, e.g., \cite{Li-2018-Information-Theoretic}):} A household's current power consumption ($d_{k-1}$) reveals real-time behaviors like appliance usage and occupancy, making its protection a core challenge in smart grids.
	\item \textbf{Building automation (see, e.g., \cite{Nekouei-2022-Optimal,Weng-2023-Optimal,Guo-2024-Privacy}):} A building's current occupancy ($d_{k-1}$) represents critical private information, making its protection a primary objective in building automation.
\end{enumerate}
\end{adjustwidth}

\begin{remark}
Our focus on protecting the latest exogenous input, $\vd_{k-1}$, at each time step $k$ stems from the high sensitivity of real-time data, such as a household’s current power consumption \cite{Li-2018-Information-Theoretic} or a building’s current occupancy \cite{Weng-2023-Optimal,Guo-2024-Privacy}, which is of primary interest to an eavesdropper.
\end{remark}

Based on the above analysis, we aim to achieve distributed fusion estimation  for the system \eqref{eq:system}, while protecting $\vd_{k-1}$ against full eavesdropping at each time step $k$.
Under privacy consideration, the linear unbiased minimum-variance state estimate given by \eqref{eq:umv:mean} cannot be transmitted to the fusion center directly as it may cause a privacy leakage (see Example 1 in \cite{Guo-2024-Privacy}). 
Therefore, a tailored privacy-preserving local state estimation needs to be designed. 

\begin{remark}\label{remark:P and d}
Note that the $\mP_{i,k|k}$ given by \eqref{eq:umv:cov} does not contain any information about $\vd_{k-1}$. Therefore, to protect $\vd_{k-1}$, we just need to modify the $\hat \vx_{i,k|k}$ given by \eqref{eq:umv:mean}.
\end{remark}

\section{Privacy-preserving local state estimation}\label{subsec:pp local estimation}

In this section, we design the privacy-preserving local state estimation based on a noise injection strategy. 
Particularly, we aim to minimize the sum of MSEs of local estimates while protecting $\vd_{k-1}$ against full eavesdropping.
To this end, it is not sufficient to consider privacy level for a single sensor; instead, a joint consideration of all the local sensors is necessary.
By augmenting all the local estimates together, denoted by
\begin{align*}
	\hat \vx_{k|k} = \big[\hat \vx_{1,k|k}^{\Trans}, \hat \vx_{2,k|k}^{\Trans}, \dots, \hat \vx_{\nS,k|k}^{\Trans}\big]^{\Trans},
\end{align*}
we know that $\hat \vx_{k|k}$ represents all the transmitted state estimates available to the full eavesdropper at time step $k$. 
 
Then, we modify $\hat \vx_{k|k}$ by injecting an independent noise as follows:
\begin{align*}
	\bar \vx_{k|k} = \hat \vx_{k|k} + \vomega_k,
\end{align*}
where $\vomega_k \sim \N(0, \mSigma_k)$ and the covariance matrix $\mSigma_k$ need to be determined.
Furthermore, to ensure that the local sensors can inject noise independently of each other, we limit the sought-after $\mSigma_k$ into the following block-diagonal form:
\begin{align*}
	\mSigma_k &= \blkdiag(\mSigma_{1,k}, \mSigma_{2,k}, \dots, \mSigma_{\nS,k}).
\end{align*}
Note that the larger $\mSigma_k$, the higher privacy level, but the lower local and fusion estimation accuracy. 
As such, an appropriate selection of $\mSigma_k$ is necessary to balance privacy level and fusion estimation accuracy.

To quantified privacy level, we adopt the commonly used ($\epsilon, \delta$)-differential privacy (see, e.g., \cite{Wei-2024-Gradient,Wang-2024-Consensus,Ny-2014-Filtering,Yan-2023-Guaranteeing,Wang-2024-Tailoring,Tie-2022-Differentially,Ye-2022-Nash,Wang-2024-Differentially}). Let $(\Omega, \mathcal F, P)$ be a probability space. Then, we first introduce the notion of differential privacy, equipped with a symmetric binary adjacency relation, denoted $\Adj(\vd_{k-1},\vd_{k-1}')$, on the space $\R^{\dd}$. 

\begin{definition}[Adjacency relation]\label{def:adj}
	Let $\epsilon_0$ be a positive real number. Then, $\Adj(\vd_{k-1},\vd_{k-1}')$ is defined as follows:
	\begin{align*}
		\Adj(\vd_{k-1},\vd_{k-1}') \Leftrightarrow \|\vd_{k-1} - \vd_{k-1}'\|_2 \leq \epsilon_0.
	\end{align*}
\end{definition}

\begin{remark}
	Definition \ref{def:adj} is commonly used in the literature (see, e.g.,  \cite{Yu-2020-Design,Yazdani-2023-Differentially}). Notably, the adjacency definition in  \cite{Han-Differentially-2016,Ny-2014-Filtering} can also be applied to our framework, denoted as $\mathcal{A}_k=\{ (\vd_{k-1}, \vd_{k-1}') \mid \vd_{k-1}, \vd_{k-1}' \in \R^{n_d}, \|\mathbf{d}_{k-1} - \mathbf{d}'_{k-1}\|_0 \le 1 \ \text{and} \ \|\mathbf{d}_{k-1} - \mathbf{d}'_{k-1}\|_\infty \le \epsilon_0 \}$. Since
		$
		\mathcal{A}_k \subset \{ (\vd_{k-1}, \vd_{k-1}') \mid \vd_{k-1}, \vd_{k-1}' \in \R^{n_d}, \|\vd_{k-1} - \vd_{k-1}'\|_2 \le \epsilon_0 \}
		$,
		Definition 2 admits a larger set of adjacent pairs than the definition well established in \cite{Han-Differentially-2016,Ny-2014-Filtering}.	
\end{remark}

\begin{definition}[Differential privacy, \cite{Ny-2014-Filtering}]\label{def:DP}
	Let ($\R^{\nS \dx},\mathcal M$) be a measurable space, and $\epsilon, \delta \geq 0$. A mechanism $M_q: \R^{\dd} \times \Omega \rightarrow \R^{\nS \dx}$ is ($\epsilon, \delta$)-differentially private if for all $\vd_{k-1}, \vd_{k-1}' \in \R^{\dd}$ such that $\Adj(\vd_{k-1}, \vd_{k-1}')$, there is
	\begin{align*}
		P(M_q(\vd_{k-1}) \in \sS) \leq e^{\epsilon}  P(M_q(\vd_{k-1}') \in \sS) + \delta, \ \forall \, \sS \in \mathcal M.
	\end{align*}
\end{definition}

\begin{remark}
	The above inequality is standard in defining the differential privacy. Since it holds for any $\vd_{k-1}, \vd_{k-1}' \in \R^{\dd}$ satisfying $\Adj(\vd_{k-1}, \vd_{k-1}')$, we can exchange $M_q(\vd_{k-1})$ with $M_q(\vd_{k-1}')$ and obtain $P(M_q(\vd_{k-1}') \in \sS) \leq e^{\epsilon}  P(M_q(\vd_{k-1}) \in \sS) + \delta$. Subtracting the two inequalities yields $1 - e^{\epsilon} - \delta \leq (1 - e^{\epsilon}) P(M_q(\vd_{k-1}') \in \sS) - \delta \leq P(M_q(\vd_{k-1}') \in \sS) - P(M_q(\vd_{k-1})  \in \sS) \leq (e^{\epsilon} - 1) P(M_q(\vd_{k-1}) + \delta \leq e^{\epsilon} - 1 + \delta$, and hence $|P(M_q(\vd_{k-1}') \in \sS) - P(M_q(\vd_{k-1})  \in \sS)| \leq e^{\epsilon} - 1 + \delta$. Since $e^{\epsilon} \approx 1 + \epsilon$ for small $\epsilon > 0$, it means that for sufficiently small $\epsilon, \delta > 0$, the eavesdropper cannot distinguish $\vd_{k-1}$ from $\vd'_{k-1}$ based on the observation $M_q$. 
	This means that $\vd_{k-1}$ is protected.
\end{remark}

Particularly in our problem, the mechanism is given as
\begin{align}\label{eq:M(d)}
	M_q(\vd_{k-1}) = \hat \vx_{k|k} + \vomega_k.
\end{align}
We next present what inequality condition does $\mSigma_{1,k}, \mSigma_{2,k}, \dots, \mSigma_{\nS,k}$ need to satisfy such that the mechanism \eqref{eq:M(d)} is ($\epsilon, \delta$)-differentially private. 
To this end, we first provide the following two lemmas.  

\begin{lemma}\label{thm:q}
The analytic expression of $\hat \vx_{k|k}$ with respect to $\vd_{k-1}$ is given as
\begin{align*}
\hat \vx_{k|k} = q(\vd_{k-1}) + \vnu_k,
\end{align*}
where
$q(\vd_{k-1}) = \mM_k \vd_{k-1} + c_k$ with
$\mM_k = \vones \otimes \mB_{k-1}$ and 
$c_k$ being a constant, and $\vnu_k$ is a zero-mean Gaussian noise with covariance matrix satisfying $\Cov(\vnu_k) \geq \mUpsilon_k$, where
$\mUpsilon_k = \bar \mG_k \mC_k \mQ_{k-1} \mC_k^{\Trans} \bar \mG_k^{\Trans}$,
$\bar \mG_k = \blkdiag(\mG_{1,k}, \mG_{2,k}, \dots, \mG_{\nS,k})$ and $	\mC_k = [\mC_{1,k}^{\Trans}, \mC_{2,k}^{\Trans}, \dots, \mC_{\nS,k}^{\Trans}]^{\Trans}$.
\end{lemma}

\textit{Proof:}
Denote
\begin{align*}
	\hat \vx_{k|k-1} &= \big[\hat \vx_{1,k|k-1}^{\Trans}, \hat \vx_{2,k|k-1}^{\Trans}, \dots, \hat \vx_{\nS,k|k-1}^{\Trans}\big]^{\Trans}, \notag\\
	\vy_k &= \big[\vy_{1,k}^{\Trans}, \vy_{2,k}^{\Trans}, \dots, \vy_{\nS,k}^{\Trans}\big]^{\Trans}, \notag\\
	\bar \mA_k &= \mI \otimes \mA_k, \notag \\
	\bar \mC_k &= \blkdiag(\mC_{1,k}, \mC_{2,k}, \dots, \mC_{\nS,k}).
\end{align*}
Then,
$\hat \vx_{k|k} = \bar \mA_{k-1} \hat \vx_{k-1|k-1} + \bar \mG_k \big(\vy_k - \bar \mC_k \bar \mA_{k-1} \hat \vx_{k-1|k-1}\big)$.
Denote $\vv_k = \big[\vv_{1,k}^{\Trans}, \vv_{2,k}^{\Trans}, \dots, \vv_{\nS,k}^{\Trans}\big]^{\Trans}$.
Then,
\begin{align*}
	\vy_k &= \mC_k \vx_k + \vv_k \notag\\
	&= \mC_k (\mA_{k-1} \vx_{k-1} + \mB_{k-1} \vd_{k-1} + \vw_{k-1}) + \vv_k \notag\\
	&= \mC_k \mA_{k-1} \vx_{k-1} + \mC_k \mB_{k-1} \vd_{k-1} + \mC_k \vw_{k-1} + \vv_k.
\end{align*}
Substituting $\vy_k$ into $\hat \vx_{k|k}$ yields
\begin{align*}
	\hat \vx_{k|k} &= \bar \mG_k \mC_k \mB_{k-1} \vd_{k-1} + \bar \mG_k \vv_k
	+ \bar \mA_{k-1} \hat \vx_{k-1 | k - 1} \notag\\
	&\quad + \bar \mG_k \bar \mC_k \bar \mA_{k-1} ( \vones \otimes \vx_{k-1} - \hat \vx_{k-1|k-1}) 
	+ \bar \mG_k \mC_k \vw_{k-1}. 
\end{align*}

Denote
\begin{align*}
	c_k &= \bar \mA_{k-1} (\vones \otimes \E[\vx_{k-1}]), \notag\\
	\vnu_k &= \bar \mA_{k-1} (\hat \vx_{k-1 | k - 1} - \vones \otimes \E[\vx_{k-1}]) + \bar \mG_k \mC_k \vw_{k-1} + \bar \mG_k \vv_k \notag \\
	&\quad + \bar \mG_k \bar \mC_k \bar \mA_{k-1} ( 1 \otimes \vx_{k-1} - \hat \vx_{k-1|k-1}). 
\end{align*}
Then,
$\hat \vx_{k|k} = \bar \mG_k \mC_k \mB_{k - 1} \vd_{k-1} + c_k + \vnu_k = \mM_k \vd_{k-1} + c_k + \vnu_k$,
where $\vnu_k$ is a zero-mean Gaussian noise with covariance matrix satisfying
$\Cov(\vnu_k) \geq \mUpsilon_k$.
$\hfill\blacksquare$

\begin{remark}
From Lemma \ref{thm:q} and \eqref{eq:M(d)}, we obtain
$
	M_q(\vd_{k-1}) = q(\vd_{k-1}) + \vnu_k + \vomega_k,
$
where $q(\vd_{k-1})$ is the so-called query function.
This indicates that the differential privacy level is collectively determined by both noise terms, $\vnu_k$ and $\vomega_k$.
\end{remark}


\begin{lemma}\label{thm:DP}
Let $\vmu_k = q(\vd_{k-1}) - q(\vd_{k-1}')$ and $\bar \mSigma_k = \Cov(\vnu_k) + \Cov(\vomega_k)$. 
Then, we have
\begin{align}\label{eq:DP}
	P(M_q(\vd_{k-1}) \in \sS) 
	&\leq e^{\epsilon} P(M_q(\vd_{k-1}') \in \sS) \notag \\
	& + \mathcal{Q}\bigg(\frac{\epsilon}{\| \vmu_k \|_{\bar \mSigma_k^{-1}}} - \frac{\| \vmu_k \|_{\bar \mSigma_k^{-1}}}{2} \bigg),
\end{align}
where $\mathcal{Q}(x) = \frac{1}{\sqrt{2 \pi}} \int_x^{\infty} \exp\{-\frac{z^2}{2}\}\diff z$ is the $\mathcal{Q}$-function.
\end{lemma}

\textit{Proof:}
For any Borel set $\sS \in \mathbb R^{\nS \dx}$, we have
\begin{align*}
	&P(M_q(\vd_{k-1}) \in \sS) \\
	&= \int_{\sS} \N(\vu; q(\vd_{k-1}), \bar \mSigma_k) \diff \vu \\
	&= \int_{\sS} (2 \pi)^{-\frac{\nS \dx}{2}} \det(\bar \mSigma_k)^{-\frac{1}{2}} \notag\\
	&\quad \cdot \exp\bigg\{-\frac{1}{2} \big\| \vu - q(\vd_{k-1}) \big\|^2_{ \bar \mSigma_k^{-1}}\bigg\} \diff \vu \\
	&= \int_{\sS} (2 \pi)^{-\frac{\nS \dx}{2}} \det( \bar \mSigma_k)^{-\frac{1}{2}} \exp\bigg\{-\frac{1}{2} \big\| \vu - q(\vd_{k-1}') \big\|^2_{\bar \mSigma_k^{-1}}\bigg\} \\
	&\quad \cdot \exp\bigg\{ \big(\vu - q(\vd_{k-1}')\big)^{\Trans} \bar \mSigma_k^{-1} \vmu_k - \frac{1}{2} \| \vmu_k \|^2_{\bar \mSigma_k^{-1}} \bigg\} \diff \vu.
\end{align*} 
Denote $f(\vu) = (\vu - q(\vd_{k-1}'))^{\Trans} \bar \mSigma_k^{-1} \vmu_k - \frac{1}{2} \| \vmu_k \|^2_{\bar \mSigma_k^{-1}}$, $\sA = \{ \vu | f(\vu) \leq \epsilon \}$. Then, we have
\begin{align*}
	&P(M_q(\vd_{k-1}) \in \sS) \\
	&= \int_{\sS \cap \sA} (2 \pi)^{-\frac{\nS \dx}{2}} \det( \bar\mSigma_k)^{-\frac{1}{2}} \exp\bigg\{-\frac{1}{2} \big\| \vu - q(\vd_{k-1}') \big\|^2_{\bar \mSigma_k^{-1}}\bigg\} \\
	&\quad \cdot \exp\{ f(\vu) \} \diff \vu + \int_{\sS \cap {\sA^c}} \N(\vu; q(\vd_{k-1}), \bar \mSigma_k) \diff \vu \\
	&\leq e^{\epsilon} P(M_q(\vd_{k-1}') \in \sS) + \int_{\sS} \N(\vu; q(\vd_{k-1}), \bar \mSigma_k) \mathcal{I}_{[f(\vu) > \epsilon]} \diff \vu,
\end{align*}
where $\sA^c$ is the complement set to $\sA$, and $\mathcal{I}_{[f(\vu) > \epsilon]}$ is an indicative function defined as
\begin{align*}
	\mathcal{I}_{[f(\vu) > \epsilon]} = \begin{cases}
		1 & f(\vu) > \epsilon \\
		0 & f(\vu) \leq \epsilon.
	\end{cases}
\end{align*}
Let $\vy = \bar \mSigma_k^{-\frac{1}{2}} (\vu - q(\vd_{k-1}))$. Then, we have
\begin{align}\label{eq:tmp2}
	&P(M_q(\vd_{k-1}) \in \sS) 
	\leq  e^{\epsilon} P(M_q(\vd_{k-1}') \in \sS) \notag \\
	& + \int_{\sS} \N(\vy; 0, \mI_{\nS \dx}) \mathcal{I}_{[\vmu_k^{\Trans} \bar \mSigma_k^{-\frac{1}{2}} \vy > -\frac{1}{2} \| \vmu_k \|^2_{\bar \mSigma_k^{-1}} + \epsilon]} \diff \vy.
\end{align}
For the right-hand side of \eqref{eq:tmp2}, we have
\begin{align*}
	&\vmu_k^{\Trans} \bar \mSigma_k^{-\frac{1}{2}} \vy > -\frac{1}{2} \| \vmu_k \|^2_{\bar \mSigma_k^{-1}} + \epsilon 
\end{align*}
which is equivalent to
\begin{align}\label{eq:tmp1}
	\Bigg\langle \frac{\vmu_k^{\Trans} \bar \mSigma_k^{-\frac{1}{2}}}{\| \vmu_k^{\Trans} \bar \mSigma_k^{-\frac{1}{2}} \|}, \vy \Bigg\rangle > -\frac{1}{2} \| \vmu_k \|_{\bar \mSigma_k^{-1}} + \frac{\epsilon}{\| \vmu_k \|_{\bar \mSigma_k^{-1}}}.
\end{align}
From \eqref{eq:tmp2} and \eqref{eq:tmp1}, we can obtain \eqref{eq:DP}.
$\hfill\blacksquare$

Lemma \ref{thm:DP} indicates that the sufficient condition to ensure ($\epsilon, \delta$)-differential privacy of the mechanism \eqref{eq:M(d)} is 
\begin{align}\label{eq:oracle condition}
\sup_{\vd_{k-1}, \vd'_{k-1} : \Adj(\vd_{k-1}, \vd'_{k-1})} \mathcal{Q}\bigg(\frac{\epsilon}{\| \vmu_k \|_{\bar \mSigma_k^{-1}}} - \frac{\| \vmu_k \|_{\bar \mSigma_k^{-1}}}{2} \bigg) \leq \delta.
\end{align}
Unfortunately, \eqref{eq:oracle condition} cannot be directly employed because the analytic computation of $\bar \mSigma_k$ is infeasible, owing to the unknown correlations among the measurement noises of the local sensors. 
To address this, we introduce an analytical lower bound of $\bar \mSigma_k$, denoted
$
\mS_k 
=\mUpsilon_k + \blkdiag(\mSigma_{1,k}, \mSigma_{2,k}, \dots, \mSigma_{\nS,k}).
$
From $\Cov(\vnu_k) \geq \mUpsilon_k$ in Lemma \ref{thm:q}, we obtain
\begin{align*}
\mS_k 
&\leq \Cov(\vnu_k)  + \blkdiag(\mSigma_{1,k}, \mSigma_{2,k}, \dots, \mSigma_{\nS,k}) \notag\\
&= \Cov(\vnu_k) + \Cov(\vomega_k) = \bar \mSigma_k.
\end{align*} 
Then, we have the following theorem.

\begin{theorem}\label{thm:DP-1}
	Consider the mechanism in \eqref{eq:M(d)}. Suppose that the injected noise covariances $\{\mSigma_{i,k}\}_{i=1}^M$ are chosen such that
	$$ \sup_{\vd_{k-1}, \vd'_{k-1} : \Adj(\vd_{k-1}, \vd'_{k-1})} \mathcal{Q}\bigg(\frac{\epsilon}{\| \vmu_k \|_{\mS_k^{-1}}} - \frac{ \| \vmu_k \|_{\mS_k^{-1}}}{2}\bigg) \le \delta.$$
	Then, the mechanism $M_q(\vd_{k-1})$ is $(\epsilon, \delta)$-differentially private.
\end{theorem}

\textit{Proof:}
From $\bar \mSigma_k \geq \mS_k$ we have
\begin{align}\label{eq:lower bound}
	&\mathcal{Q}\bigg(\frac{\epsilon}{\| \vmu_k \|_{\bar \mSigma_k^{-1}}} - \frac{\| \vmu_k \|_{\bar \mSigma_k^{-1}}}{2} \bigg) 
	\leq \mathcal{Q}\bigg(\frac{\epsilon}{\| \vmu_k \|_{\mS_k^{-1}}} - \frac{ \| \vmu_k \|_{\mS_k^{-1}}}{2}\bigg).
\end{align}
Then, it follows from Lemma \ref{thm:DP} and Definition \ref{def:DP} that $M_q(\vd_{k-1})$ is ($\epsilon, \delta$)-differentially  private. 
$\hfill\blacksquare$

\begin{remark}
For any pair of $\vd_{k-1}$ and $\vd_{k-1}'$ satisfying $\Adj(\vd_{k-1}, \vd'_{k-1})$, Theorem \ref{thm:DP-1} requires that the difference $\vmu_k = q(\vd_{k-1}) - q(\vd_{k-1}')$ satisfies the inequality:
	$\frac{\epsilon}{\| \vmu_k \|_{\mS_k^{-1}}} - \frac{ \| \vmu_k \|_{\mS_k^{-1}} }{2} \ge \mathcal{Q}^{-1}(\delta)$.	
	A smaller Mahalanobis distance $\| \vmu_k \|_{\mS_k^{-1}}$ indicates that $q(\vd_{k-1})$ and $q(\vd_{k-1}')$ are statistically closer and thus harder to distinguish, thereby enhancing privacy. 
	Thus,  one may reduce $\| \vmu_k \|_{\mS_k^{-1}}$ by increasing the covariance of the injected noise. This enlarges $\mS_k$, thereby reducing $\| \vmu_k \|_{\mS_k^{-1}}$ since $\mS_k^{-1}$ becomes smaller. 
	Observe that the function
	$f(x) = \frac{\epsilon}{x} - \frac{x}{2}$
	is decreasing for $x > 0$. Therefore, reducing $\| \vmu_k \|_{\mS_k^{-1}}$ increases the value of $f\left(\| \vmu_k \|_{\mS_k^{-1}}\right)$, making Theorem \ref{thm:DP-1}'s condition easier to meet. Consequently, a greater noise covariance leads to a higher $(\epsilon, \delta)$-differential privacy level.
\end{remark}

Based on Theorem \ref{thm:DP-1}, we construct the following constrained minimization problem, aiming at minimizing the sum of MSEs of the local estimates while ensuring ($\epsilon, \delta$)-differential privacy:
\begin{equation}\label{min:original}
	\begin{aligned}
		\min & \qquad \sum_{i = 1}^{\nS}\trace(\mSigma_{i,k}) \\
		\mathrm{s.t.} & \sup_{\vd_{k-1}, \vd'_{k-1} : \Adj(\vd_{k-1}, \vd'_{k-1})} \mathcal{Q}\bigg(\frac{\epsilon}{\| \vmu_k \|_{\mS_k^{-1}}} - \frac{ \| \vmu_k \|_{\mS_k^{-1}}}{2}\bigg) \leq \delta \\
		&\ \blkdiag(\mSigma_{1,k}, \mSigma_{2,k}, \dots, \mSigma_{\nS,k}) \geq 0.
	\end{aligned}
\end{equation}

\begin{remark}
The first constraint of \eqref{min:original} is a joint constraint on the entire block-diagonal noise covariance matrix $\mSigma_k = \blkdiag(\mSigma_{1,k}, ..., \mSigma_{M,k})$.
The challenge, therefore, is to optimally allocate the privacy budget (in the form of noise covariance) among the different local sensors.
Additionally, the optimization variables $\mSigma_{1,k}, \mSigma_{2,k}, \dots, \mSigma_{\nS,k}$ are not restricted to being isotropic (i.e., $\mSigma_{i,k} = \sigma_i \mI_{\dx}$, $i = 1, 2, \dots, \nS$). 
This is more general than most of the existing works where the parameters to be determined are scalars (see, e.g., \cite{Wang-2024-Consensus,Ny-2014-Filtering,Yan-2023-Guaranteeing,Tie-2022-Differentially,Ye-2022-Nash}).
However, this increased generality also results in a greater complexity for the minimization problem.
\end{remark}

\begin{remark}
		It follows from \eqref{eq:lower bound} that the use of $\mS_k$ relaxes the privacy constraint in  \eqref{min:original}, but comes at the cost of increased noise injection. This leads to a larger state estimation covariance, demonstrating a direct trade-off between privacy level and estimation accuracy.
\end{remark}

On the one hand, it is not difficult to verify that the first constraint is non-convex. Thus, the problem \eqref{min:original} is non-convex and its analytic solution is hard to obtain.
On the other hand, computational efficiency is critically important in real-time state estimation. Thus, it is of great significance to develop an efficient algorithm to solve the problem \eqref{min:original}. 
To this end, an approach of relaxation is proposed, as presented in the following theorem.

\begin{theorem}[SDP relaxation for optimal noise design]
	Let $b = \frac{\epsilon_0^2 \| \mM_k \|^2_2}{(-\mathcal{Q}^{-1}(\delta) + \sqrt{(\mathcal{Q}^{-1}(\delta))^2 + 2 \epsilon})^2}$.
	Then, the original problem \eqref{min:original} can be relaxed to the following SDP problem:
	\begin{equation}\label{min:SDP}
		\begin{aligned}
			\min & \qquad \sum_{i = 1}^{\nS}\trace(\mSigma_{i,k}) \\
			\mathrm{s.t.} & \qquad \blkdiag(\mSigma_{1,k}, \mSigma_{2,k}, \dots, \mSigma_{\nS,k}) \\ & \qquad + \mUpsilon_k - b \mI_{\nS \dx} \geq 0, \\
			&\qquad \blkdiag(\mSigma_{1,k}, \mSigma_{2,k}, \dots, \mSigma_{\nS,k}) \geq 0.
		\end{aligned}
	\end{equation}
\end{theorem}

\textit{Proof:}
Due to
\begin{align*}
	&\sup_{q,q'} \mathcal{Q}\bigg(-\frac{1}{2} \| \vmu_k \|_{\mS_k^{-1}} + \frac{\epsilon}{\| \vmu_k \|_{\mS_k^{-1}}}\bigg) \notag\\
	&=  \mathcal{Q}\Bigg(\inf_{q,q'} \bigg(-\frac{1}{2} \| \vmu_k \|_{\mS_k^{-1}} + \frac{\epsilon}{\| \vmu_k \|_{\mS_k^{-1}}}\bigg)\Bigg),
\end{align*}
the first constraint of \eqref{min:original} is equivalent to
\begin{align*}
	&\mathcal{Q}\Bigg(\inf_{q,q'} \bigg(-\frac{1}{2} \| \vmu_k \|_{\mS_k^{-1}} + \frac{\epsilon}{\| \vmu_k \|_{\mS_k^{-1}}}\bigg)\Bigg) \leq \delta, 
\end{align*}
and further equivalent to
\begin{align*}
	\inf_{q,q'} \bigg(-\frac{1}{2} \| \vmu_k \|_{\mS_k^{-1}} + \frac{\epsilon}{\| \vmu_k \|_{\mS_k^{-1}}}\bigg) \geq \mathcal{Q}^{-1}(\delta).
\end{align*}
Since $f(x) = -\frac{1}{2}x + \frac{\epsilon}{x}$ is a monotonically decreasing function, we should find $\sup_{q,q'}\| \vmu_k \|_{\mS_k^{-1}}$ to obtain $\inf_{q,q'} (-\frac{1}{2} \| \vmu_k \|_{\mS_k^{-1}} + \frac{\epsilon}{\| \vmu_k \|_{\mS_k^{-1}}})$. Due to
\begin{align*}
	&\sup_{q,q'}\| \vmu_k \|_{\mS_k^{-1}}^2 \\
	&=\sup_{\vd_{k-1},\vd_{k-1}'} \big((\vd_{k-1} - \vd_{k-1}')^{\Trans} \mM_k^{\Trans} \mS_k^{-1} \mM_k (\vd_{k-1} - \vd_{k-1}')\big) \\
	&= \| \mM_k^{\Trans} \mS_k^{-1} \mM_k \|_2 \epsilon_0^2,
\end{align*}
where $ \| \mM_k^{\Trans} \mS_k^{-1} \mM_k \|_2$ represents the maximum singular value of $\mM_k^{\Trans} \mS_k^{-1} \mM_k$,
the first constraint of \eqref{min:original} is further equivalent to
\begin{align*}
	-\frac{1}{2}  \| \mM_k^{\Trans} \mS_k^{-1} \mM_k \|_2^{\frac{1}{2}} \epsilon_0 + \frac{\epsilon}{ \| \mM_k^{\Trans} \mS_k^{-1} \mM_k \|_2^{\frac{1}{2}} \epsilon_0} \geq \mathcal{Q}^{-1}(\delta).
\end{align*}
Denote $\| \mM_k^{\Trans} \mS_k^{-1} \mM_k \|_2 = a$.
Then, we have
\begin{align*}
	&-\frac{1}{2} \epsilon_0 \sqrt{a} + \frac{\epsilon}{\epsilon_0 \sqrt{a}} \geq \mathcal{Q}^{-1}(\delta), 
\end{align*}
which is equivalent to
\begin{align*}
	-\frac{1}{2} \epsilon_0 a - \mathcal{Q}^{-1}(\delta) \sqrt{a} + \frac{\epsilon}{\epsilon_0} \geq 0, \ a \geq 0.
\end{align*}
Thus, we can obtain
\begin{align*}
	 0 \leq a \leq \frac{(-\mathcal{Q}^{-1}(\delta) + \sqrt{(\mathcal{Q}^{-1}(\delta))^2 + 2 \epsilon})^2}{\epsilon_0^2}.
\end{align*}
Then, we relax the first constraint of \eqref{min:original}  to
\begin{align*}
	\| \mM_k^{\Trans} \mS_k^{-1} \mM_k \|_2 &\leq \| \mM_k \|^2_2 \| \mS_k^{-1} \|_2 \notag\\
	&\leq \frac{(-\mathcal{Q}^{-1}(\delta) + \sqrt{(\mathcal{Q}^{-1}(\delta))^2 + 2 \epsilon})^2}{\epsilon_0^2},
\end{align*}
and thus,
\begin{align*}
	 \| \mS_k^{-1} \|_2 
	&\leq \frac{(-\mathcal{Q}^{-1}(\delta) + \sqrt{(\mathcal{Q}^{-1}(\delta))^2 + 2 \epsilon})^2}{\epsilon_0^2 \| \mM_k \|^2_2}, 
\end{align*}
which is equivalent to
\begin{align*}
	 \lambda_{\min}(\mS_k)
	&\geq \frac{\epsilon_0^2 \| \mM_k \|^2_2}{(-\mathcal{Q}^{-1}(\delta) + \sqrt{(\mathcal{Q}^{-1}(\delta))^2 + 2 \epsilon})^2}.
\end{align*}
Denoting $b = \frac{\epsilon_0^2 \| \mM_k \|^2_2}{(-\mathcal{Q}^{-1}(\delta) + \sqrt{(\mathcal{Q}^{-1}(\delta))^2 + 2 \epsilon})^2}$, the problem \eqref{min:original} can be relaxed to the SDP problem \eqref{min:SDP}. $\hfill\blacksquare$


\begin{remark}
The SDP relaxation \eqref{min:SDP} offers a twofold benefit. First, it is computationally tractable and can be efficiently solved by standard packages like CVX (see \cite{cvx}). Second, and crucially, any feasible solution to \eqref{min:SDP} also satisfies the constraints of the original problem \eqref{min:original}. This implies that the ($\epsilon, \delta$)-differential  privacy of \eqref{eq:M(d)} is maintained. Thus, this relaxation yields a practical solution method without compromising the privacy guarantee.
\end{remark}

To assess the effectiveness of the SDP relaxation \eqref{min:SDP}, the following proposition quantifies the relaxation quality by providing an upper bound on the gap between the SDP relaxation \eqref{min:SDP} and the original problem \eqref{min:original}.

\begin{proposition}[Upper bound on the relaxation gap]\label{prop:relaxation gap}
	Let $J_{\text{orig}}^*$ and $J_{\text{relax}}^*$ be the minimums of the original problem \eqref{min:original} and the SDP problem \eqref{min:SDP}, respectively, and $\mS_{k, \text{orig}}^* = \mUpsilon_k + \mSigma_{k, \text{orig}}^*$ with $\mSigma_{k, \text{orig}}^*$ being the optimal solution of \eqref{min:original}. Then, the relaxation gap is bounded by
	\begin{align*}
	J_{\text{relax}}^* - J_{\text{orig}}^* \le \left( \max\left(1, \frac{b}{\lambda_{\rm{min}}(\mS_{k, \text{orig}}^*)}\right) - 1 \right) \trace(\mS_{k, \text{orig}}^*).
	\end{align*} 
  
\end{proposition}
	
\textit{Proof:}
	The feasible set of the SDP is a subset of the feasible set of the original problem, implying
	$J_{\text{relax}}^*  \ge J_{\text{orig}}^*$.
	To provide an upper bound on the gap $J_{\text{relax}}^* - J_{\text{orig}}^*$, we construct a feasible solution for the SDP based on the original optimal solution $\mSigma_{k, \text{orig}}^*$. Define the scaling factor
	\begin{align*} 
	\alpha^* = \max(1, b / \lambda_{\rm{min}}(\mS_{k, \text{orig}}^*)). 
	\end{align*}
	Then, from $\alpha^* \ge 1$ and $\mS_{k, \text{orig}}^* > 0$, the scaled matrix $\mS_k' = \alpha^* \mS_{k, \text{orig}}^*$ is positive definite. Furthermore,
	\begin{align*}
	\lambda_{\rm{min}}(\mS_k') &= \alpha^* \lambda_{\rm{min}}(\mS_{k, \text{orig}}^*) \\
	&\ge (b / \lambda_{\rm{min}}(\mS_{k, \text{orig}}^*)) \lambda_{\rm{min}}(\mS_{k, \text{orig}}^*) \\
	&= b. 
	\end{align*}
	Thus, $\mS_k' \ge b\mI_{\nS \dx}$, which means $\mS_k'$ satisfies the SDP constraint. 
	The corresponding injected noise is
	\begin{align*}
	\mSigma_k' &= \mS_k' - \mUpsilon_k \\
	&= \alpha^* \mS_{k, \text{orig}}^* - \mUpsilon_k \\
	&= \alpha^*(\mUpsilon_k + \mSigma_{k, \text{orig}}^*) - \mUpsilon_k \\
	&= \alpha^* \mSigma_{k, \text{orig}}^* + (\alpha^*-1)\mUpsilon_k. 
	\end{align*}
	Since $\alpha^* \ge 1$, $\mSigma_{k, \text{orig}}^* \ge 0$, and $\mUpsilon_k \ge 0$, we have $\mSigma_k' \ge 0$. Therefore, $\mSigma_k'$ is a feasible solution for the SDP problem \eqref{min:SDP}.
	
	Since $J_{\text{relax}}^*$ is the minimum for the SDP, it must be less than or equal to the objective value for any feasible solution, including $\mSigma_k'$:
	\begin{align*}
	J_{\text{relax}}^* &\le \trace(\mSigma_k') \\
	&= \trace(\alpha^* \mSigma_{k, \text{orig}}^* + (\alpha^*-1)\mUpsilon_k) \\
	&= \alpha^* J_{\text{orig}}^* + (\alpha^*-1)\trace(\mUpsilon_k).
	\end{align*}
	Then, the gap between the minimums can be bounded as:
	\begin{align*} J_{\text{relax}}^* - J_{\text{orig}}^* &\le (\alpha^* - 1) J_{\text{orig}}^* + (\alpha^*-1)\trace(\mUpsilon_k) \\ &= (\alpha^* - 1) (\trace(\mSigma_{k, \text{orig}}^*) + \trace(\mUpsilon_k)) \\ &= (\alpha^* - 1) \trace(\mS_{k, \text{orig}}^*). \end{align*}
	Substituting $\alpha^*$ into the above inequality implies:
	$$ J_{\text{relax}}^* - J_{\text{orig}}^* \le \left( \max\left(1, \frac{b}{\lambda_{\rm{min}}(\mS_{k, \text{orig}}^*)}\right) - 1 \right) \trace(\mS_{k, \text{orig}}^*). $$
	This completes the proof. 
$\hfill\blacksquare$

According to Proposition \ref{prop:relaxation gap}, the relaxation gap vanishes if $\lambda_{\rm{min}}(\mS_{k, \text{orig}}^*) \geq b$ (the relaxation is tight); otherwise, its upper bound monotonically increases with the ratio  $b / \lambda_{\rm{min}}(\mS_{k, \text{orig}}^*)$.

\section{Privacy-preserving distributed fusion estimation}

At the fusion center side, after receiving all the noisy state estimates and error covariance matrices, the fusion estimate, denoted by $\hat \vx_{k|k}^{\cen}$, and the corresponding error covariance matrix, denoted by $\mP_{k|k}$, are derived by utilizing the approach of covariance intersection \cite{Julier-1997-A}:
\begin{align}
	\mP_{k|k}^{-1} \hat \vx_{k|k}^{\cen} &= \sum_{i = 1}^{\nS} w_i \bar \mP_{i,k|k}^{-1} \bar \vx_{i,k|k}, \label{eq:CI:mean}\\
	\mP_{k|k}^{-1} &= \sum_{i = 1}^{\nS} w_i \bar \mP_{i,k|k}^{-1}, \label{eq:CI:cov}
\end{align}
where $\bar \vx_{i,k|k} = \hat \vx_{i,k|k} + \vomega_{i,k}$, $\vomega_{i,k} \sim \N(0, \mSigma_{i,k}^*)$, and $\bar \mP_{i,k|k} = \mP_{i,k|k} + \mSigma_{i,k}^*$.


The proposed differentially private distributed fusion estimation algorithm is summarized in Algorithm \ref{alg:PP-ACF}. 
It should be noted in Algorithm \ref{alg:PP-ACF} that the local sensors uses the optimal local state estimates given by \eqref{eq:umv:mean} for the next prediction step. The noisy local estimates are adopted at the fusion center only. Besides, the process that the fusion center sends $\{\mSigma_{i,k}^*\}_{i = 1}^{\nS}$ to the local sensors will not cause a privacy leakage since the full eavesdropper cannot obtain the sampled values of the injected noises from $\{\mSigma_{i,k}^*\}_{i = 1}^{\nS}$.


\begin{algorithm}[htbp]
	\renewcommand{\algorithmicrequire}{\textbf{Input:}}
	\renewcommand{\algorithmicensure}{\textbf{Output:}}
	\caption{Differentially private distributed fusion estimation algorithm}
	\label{alg:PP-ACF}
	\begin{algorithmic}[1]
		\REQUIRE{$\hat \vx_{i,k-1|k-1}$, $\mP_{i,k-1|k-1}$}
		\FOR{$i = 1, 2,\dots, \nS$}
		\STATE Node $i$ calculates $\hat \vx_{i,k|k-1}$ and $\mP_{i,k|k-1}$ using \eqref{eq:non-privacy:pre:mean} and \eqref{eq:non-privacy:pre:cov}.
	    \STATE Node $i$ calculates  $\hat \vx_{i,k|k}$ and $\mP_{i,k|k}$ using \eqref{eq:umv:mean} and \eqref{eq:umv:cov}.
	    \ENDFOR
	    \STATE Fusion center solves \eqref{min:SDP} to get
	    \begin{align*}
	    	\mSigma_k^* = \blkdiag(\mSigma_{1,k}^*, \mSigma_{2,k}^*, \dots, \mSigma_{\nS,k}^*),
	    \end{align*}
	    and sends $\mSigma_{i,k}^*$ to Node $i$.
	    \FOR{$i = 1, 2,\dots, \nS$}
	    \STATE Node $i$ generates $\vomega_{i,k} \sim \N(0, \mSigma_{i,k}^*)$.
	    \STATE Node $i$ calculates $\bar \vx_{i,k|k} = \hat \vx_{i,k|k} + \vomega_{i,k}$.
	    \STATE Node $i$ calculates $\bar \mP_{i,k|k} = \mP_{i,k|k} + \mSigma_{i,k}^*$.
	    \STATE Node $i$ sends $\bar \vx_{i,k|k}$ and $\bar \mP_{i,k|k}$ to the fusion center.
	    \ENDFOR
	    \STATE Fusion center calculates $\hat \vx_{k|k}^{\cen}$ and $ \mP_{k|k}$ using \eqref{eq:CI:mean} and \eqref{eq:CI:cov}.
	    \ENSURE{$\hat \vx_{k|k}^{\cen}$, $\mP_{k|k}$}
	\end{algorithmic}
\end{algorithm}

\begin{remark}\label{rem:CI consistency}
	The covariance intersection fusion yields a consistent estimate, that is, the fused covariance matrix $\mP_{k|k}$ satisfies $\E[(\hat \vx_{k|k}^{\cen} - \vx_k)(\hat \vx_{k|k}^{\cen} - \vx_k)^{\Trans}] \leq \mP_{k|k}$, conservatively bounding the true error to prevent over-confidence and filter divergence. This holds provided that the covariance matrix from each local sensor upper-bounds its true error \cite{Julier-1997-A}. 	
	In Algorithm \ref{alg:PP-ACF}, the fusion center receives the noisy state estimate $\bar{\vx}_{i,k|k} = \hat{\vx}_{i,k|k} + \vomega_{i,k}$, whose true error covariance matrix is 
	$
	\mathbb{E}\left[ (\bar{\vx}_{i,k|k} - \vx_k)(\bar{\vx}_{i,k|k} - \vx_k)^{\Trans} \right] = \mP_{i,k|k} + \mSigma_{i,k}^* = \bar{\mP}_{i,k|k}.
	$
	Therefore, although $\mP_{i,k|k}$ contains no information about $\vd_{k-1}$ (Remark \ref{remark:P and d}), transmitting the augmented covariance matrix $\bar{\mP}_{i,k|k}$ is required for consistency.
\end{remark}


	The following proposition quantifies the estimation accuracy loss incurred by privacy protection. The loss, defined as $\Delta \mP_{k|k} = \mP_{k|k} - \mP_{k|k}^{\text{non-priv}} \ge 0$ at time step $k$, is derived by comparing the error covariance matrix of Algorithm \ref{alg:PP-ACF} with that of the non-private, linear unbiased minimum-variance estimator.  

\begin{proposition}[Estimation accuracy loss]\label{prop:perfor:loss}
	The estimation accuracy loss of Algorithm \ref{alg:PP-ACF} at time step $k$ is given as:
	\begin{align*}
		\Delta {\mP}_{k|k} &= \mP_{k|k} \left( \sum_{i=1}^{M} w_i \mP_{i,k|k}^{-1}\mSigma_{i,k}^*(\mP_{i,k|k} + \mSigma_{i,k}^*)^{-1} \right) \\
		&\quad \ \cdot {\mP}_{k|k}^{\text{non-priv}}.
	\end{align*}
\end{proposition}

\textit{Proof:}
	From \eqref{eq:CI:cov} and the definition of $\mP_{k|k}^{\text{non-priv}}$, it follows that
	\begin{align*}
	&(\mP_{k|k}^{\text{non-priv}})^{-1} - \mP_{k|k}^{-1} \\
	&= \sum_{i=1}^{\nS} w_i \big( \mP_{i,k|k}^{-1} - (\mP_{i,k|k} + \mSigma_{i,k}^*)^{-1} \big) \\
	&= \sum_{i=1}^{\nS} w_i \big( \mP_{i,k|k}^{-1} (\mP_{i,k|k} + \mSigma_{i,k}^*) (\mP_{i,k|k} + \mSigma_{i,k}^*)^{-1} \\
	&\quad - (\mP_{i,k|k} + \mSigma_{i,k}^*)^{-1} \big) \\
	&= \sum_{i=1}^{\nS} w_i \big( (\mI_{\dx} + \mP_{i,k|k}^{-1}  \mSigma_{i,k}^*) - \mI_{\dx} \big) (\mP_{i,k|k} + \mSigma_{i,k}^*)^{-1} \\
	&= \sum_{i=1}^{\nS} w_i \mP_{i,k|k}^{-1}\mSigma_{i,k}^*(\mP_{i,k|k} + \mSigma_{i,k}^*)^{-1}. 
\end{align*}
Substituting the above equality into $\Delta \mP_{k|k}$, it follows that:
	\begin{align*}
		\Delta \mP_{k|k} &= \mP_{k|k} - \mP_{k|k}^{\text{non-priv}} \\
		&= \mP_{k|k} \big((\mP_{k|k}^{\text{non-priv}})^{-1} - \mP_{k|k}^{-1} \big) \mP_{k|k}^{\text{non-priv}} \\
		&= \mP_{k|k} \Bigg( \sum_{i=1}^{M} w_i \mP_{i,k|k}^{-1}\mSigma_{i,k}^*(\mP_{i,k|k} + \mSigma_{i,k}^*)^{-1} \Bigg) \\
		&\quad \ \cdot \mP_{k|k}^{\text{non-priv}}.
	\end{align*}
This completes the proof.
$\hfill\blacksquare$

\begin{remark}
	Based on Proposition~\ref{prop:perfor:loss}, we have the following observations:
	i) Under small injected noise (i.e., $\|\mSigma_{i,k}^*\|$ is small relative to $\|\mP_{i,k|k}\|$), the inverse term admits the approximation $(\mP_{i,k|k} + \mSigma_{i,k}^*)^{-1} \approx \mP_{i,k|k}^{-1} - \mP_{i,k|k}^{-1}\mSigma_{i,k}^* \mP_{i,k|k}^{-1}$, and thus ${\mP}_{k|k}^{\text{priv}} \approx {\mP}_{k|k}^{\text{non-priv}}$. Consequently, the estimation accuracy loss can be approximated as
	$\Delta {\mP}_{k|k} \approx {\mP}_{k|k}^{\text{non-priv}} \left( \sum_{i=1}^{M} w_i \mP_{i,k|k}^{-1} \mSigma_{i,k}^* \mP_{i,k|k}^{-1} \right) {\mP}_{k|k}^{\text{non-priv}}$,
	which indicates that $\Delta {\mP}_{k|k}$ is approximately linear in $\mSigma_{k}^*$.
	ii) The result also reveals a trade-off between privacy level and estimation accuracy: a higher privacy level implies a larger noise covariance  $\mSigma_k^*$, which in turn enlarges the  estimation accuracy loss $\Delta \mP_{k|k}$.
\end{remark}

\begin{figure}[htbp]
	\centering
	\includegraphics[width=0.44\textwidth]{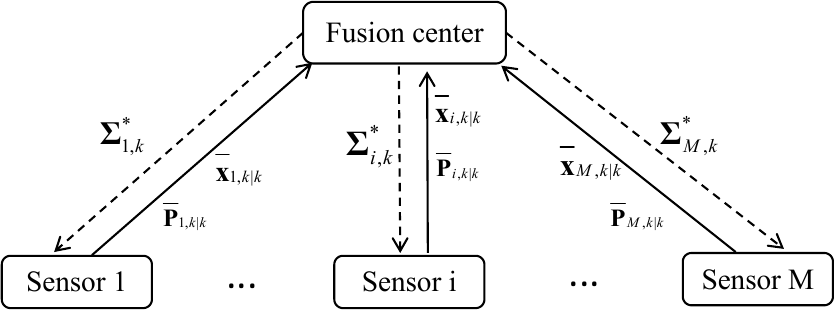}
	\caption{Information transmission in Algorithm \ref{alg:PP-ACF}}
	\label{fig:without feedback}
\end{figure}

The information transmission in Algorithm \ref{alg:PP-ACF} is illustrated in Fig.~\ref{fig:without feedback}. 
We can see from Fig.~\ref{fig:without feedback} that the local sensors send their state estimates $\{\bar \vx_{i,k|k}\}_{i = 1}^{\nS}$ and error covariance matrices $\{\bar \mP_{i,k|k}\}_{i = 1}^{\nS}$ to the fusion center, 
but the fusion center does not send the fusion estimate $\hat \vx_{k|k}^{\cen}$ and error covariance matrix $\mP_{k|k}$ back to the local sensors.

Accuracy is crucial in distributed fusion estimation as it directly improves the reliability of decision-making in various applications. 
To enhance the fusion estimation accuracy of Algorithm \ref{alg:PP-ACF} while ensuring the same ($\epsilon, \delta$)-differential privacy, 
we introduce a feedback mechanism. 
Specifically, after implementing Algorithm \ref{alg:PP-ACF}, the fusion center further sends the fusion estimate $\hat \vx_{k|k}^{\cen}$ and error covariance matrix $\mP_{k|k}$ back to all the local sensors. 
Then, each node $i$ updates its local estimate and error covariance matrix to $\hat \vx_{i,k|k}^{\pre}$ and $\mP_{i,k|k}^{\pre}$, by fusing its local estimate $\hat \vx_{i,k|k}$ and error covariance matrix $\mP_{i,k|k}$ with the fusion estimate $\hat \vx_{k|k}^{\cen}$ and error covariance matrix $\mP_{k|k}$. 
The differentially private distributed fusion estimation algorithm with enhanced accuracy via a feedback mechanism is summarized in Algorithm \ref{alg:PP-ACF:feedback}.

\begin{algorithm}[htbp]
	\renewcommand{\algorithmicrequire}{\textbf{Input:}}
	\renewcommand{\algorithmicensure}{\textbf{Output:}}
	\caption{Differentially private distributed fusion estimation algorithm with enhanced accuracy via a feedback mechanism}
	\label{alg:PP-ACF:feedback}
	\begin{algorithmic}[1]
		\REQUIRE{$\hat \vx_{i,k-1|k-1}^{\pre}$, $\mP_{i,k-1|k-1}^{\pre}$}
		\STATE Implement Algorithm \ref{alg:PP-ACF} to get $\hat \vx_{k|k}^{\cen, \update}$ and $ \mP_{k|k}^{\update}$ at the fusion center, where $(\hat \vx_{i,k-1|k-1}, \mP_{i,k-1|k-1})$ is replaced by $(\hat \vx_{i,k-1|k-1}^{\pre} , \mP_{i,k-1|k-1}^{\pre})$.
		\STATE Fusion center sends $\hat \vx_{k|k}^{\cen, \update}$ and $ \mP_{k|k}^{\update}$ to all the local sensors.
		\FOR{$i = 1, 2, \dots, \nS$}
		\STATE Node $i$ updates its local estimate and error covariance matrix using the approach of covariance intersection:
		\begin{align*}
			\big(\mP_{i,k|k}^{\pre}\big)^{-1} \hat \vx_{i,k|k}^{\pre} &= v_1^{(i)} \big(\mP_{i,k|k}^{\update}\big)^{-1} \hat \vx_{i,k|k}^{\update} \\
			&\quad + v_2^{(i)} \big( \mP_{k|k}^{\update}\big)^{-1} \hat \vx_{k|k}^{\cen, \update}, \\
			\big(\mP_{i,k|k}^{\pre}\big)^{-1} &= v_1^{(i)} \big(\mP_{i,k|k}^{\update}\big)^{-1} \\
			&\quad + v_2^{(i)} \big( \mP_{k|k}^{\update}\big)^{-1},
		\end{align*}
	    where the weights $v_1^{(i)}$,  $v_2^{(i)}$ are chosen as follows:
    \begin{itemize}
    \item If $\mP_{k|k}^{\update} \le \mP_{i,k|k}^{\update}$, set $v_1^{(i)} = 0, v_2^{(i)} = 1$;
    \item Otherwise, set $v_1^{(i)} = 1, v_2^{(i)} = 0$.
    \end{itemize}
		\ENDFOR
		\ENSURE{$\hat \vx_{k|k}^{\cen, \update}$, $ \mP_{k|k}^{\update}$}
	\end{algorithmic}
\end{algorithm}

\begin{remark}
	By implementing Step 4 and Step 5 in Algorithm \ref{alg:PP-ACF:feedback}, each node enhances its local estimation accuracy at time step $k$. Furthermore, the fusion estimation accuracy for the next time step $k + 1$ will be enhanced. Despite an enhanced accuracy, there is always a loss due to the noise injection strategy, which reflects the trade-off between differential privacy level and fusion estimation accuracy.
\end{remark}

\begin{figure}[htbp]
	\centering
	\includegraphics[width=0.48\textwidth]{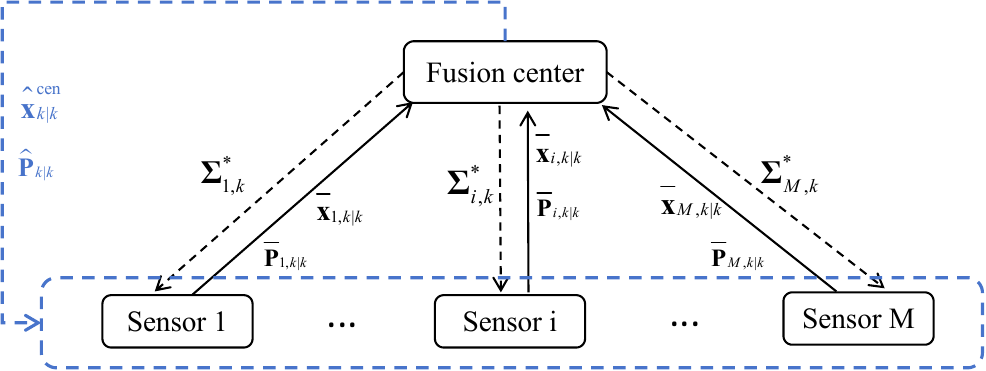}
	\caption{Information transmission in Algorithm \ref{alg:PP-ACF:feedback}}
	\label{fig:with feedback}
\end{figure}


The information transmission in Algorithm \ref{alg:PP-ACF:feedback} is illustrated in Fig.~\ref{fig:with feedback}. 
We can see from Figs.~\ref{fig:without feedback} and \ref{fig:with feedback} that Algorithm \ref{alg:PP-ACF:feedback} needs to transmit $\hat \vx_{k|k}^{\cen}$ and $ \mP_{k|k}$, but Algorithm \ref{alg:PP-ACF} does not. This implies that the full eavesdropper can acquire more data in Algorithm \ref{alg:PP-ACF:feedback}.
Then, a natural question is whether Algorithm \ref{alg:PP-ACF:feedback} can ensure the same ($\epsilon, \delta$)-differential privacy while achieving enhanced estimation accuracy over Algorithm \ref{alg:PP-ACF}?
The following two propositions give the answer.

\begin{proposition}\label{prop:DP}
Algorithm \ref{alg:PP-ACF:feedback} ensures the same ($\epsilon,\delta$)-differential privacy as Algorithm \ref{alg:PP-ACF}.
\end{proposition}

\textit{Proof:} 
	For Algorithm \ref{alg:PP-ACF:feedback}, denote
	\begin{align*}
	\mathcal{X}_k=\{\bar \vx_{1,k|k}, \bar \vx_{2,k|k}, \dots, \bar \vx_{\nS,k|k}, \hat \vx_{k|k}^{\cen}\}.
	\end{align*} 
	Then, $\mathcal{X}_k$ represents all the state estimates available to the full eavesdropper for inferring $\vd_{k-1}$ at time step $k$. 
    According to Algorithm \ref{alg:PP-ACF}, the estimates $\bar \vx_{1,k|k}, \bar \vx_{2,k|k}, \dots, \bar \vx_{\nS,k|k}$ satisfy ($\epsilon, \delta$)-differential privacy. From \eqref{eq:CI:mean}, it follows that
\begin{align*}
\hat \vx_{k|k}^{\cen} = \sum_{i = 1}^{\nS} w_i \mP_{k|k} \mP_{i,k|k}^{-1} \bar \vx_{i,k|k}.
\end{align*}
Thus, $\hat \vx_{k|k}^{\cen}$ is a linear combination of the state estimates $\bar \vx_{1,k|k}, \bar \vx_{2,k|k}, \dots, \bar \vx_{\nS,k|k}$. Furthermore, from the resilience of differential privacy to post-processing (see, e.g., Theorem 1 of \cite{Ny-2014-Filtering}), $\hat \vx_{k|k}^{\cen}$ satisfies ($\epsilon, \delta$)-differential privacy. Therefore, the set $\mathcal{X}_k$ also satisfies ($\epsilon, \delta$)-differential privacy.
$\hfill\blacksquare$


\begin{proposition}\label{prop:accuracy_enhancement}
	Algorithm \ref{alg:PP-ACF:feedback} enhances the fusion and local estimation accuracy of Algorithm \ref{alg:PP-ACF}, i.e.,
	\begin{align*}
	\mP_{k|k}^{\update} \leq \mP_{k|k}, \
	\mP_{i,k|k}^{\pre} \leq \mP_{i,k|k}, \ \text{for all} \ k.
	\end{align*}
\end{proposition}

\textit{Proof:}
	The proof proceeds by mathematical induction.
	
	\textbf{Base step ($k=1$).}
	Up to the fusion estimation at the fusion center, both algorithms are identical:
	\begin{align*}
	\mP_{i,1|1} = \mP_{i,1|1}^{\update},\  \mP_{1|1} = \mP_{1|1}^{\update}.
	\end{align*} 
	Then, Algorithm~\ref{alg:PP-ACF} remains $\mP_{i,1|1}$, while
	Algorithm~\ref{alg:PP-ACF:feedback} updates $\mP_{i,1|1}^{\update}$ via Step 4 and Step 5.
	The local update strategy ensures 
	\begin{align*}
	\mP_{i,1|1}^{\pre} \le \mP_{i,1|1}^{\update} = \mP_{i,1|1}, \ i = 1, 2, \dots, \nS.
	\end{align*}

	\textbf{Inductive step.}
	Assume that $\mP_{i,k-1|k-1}^{\pre} \le \mP_{i,k-1|k-1}$ and $\mP_{k-1|k-1}^{\update} \le \mP_{k-1|k-1}$. Then, we prove $\mP_{i,k|k}^{\pre} \le \mP_{i,k|k}$ and $\mP_{k|k}^{\update} \le \mP_{k|k}$ as follows.
	
	Following the map $\kappa_{k-1}: \mP_{i,k-1|k-1} \mapsto \mP_{i,k|k-1}$ given by \eqref{eq:non-privacy:pre:cov} is operator monotone,  $\mP_{i,k-1|k-1}^{\update} \le \mP_{i,k-1|k-1}$ indicates
	\begin{align*}
	\mP_{i,k|k-1}^{\update} \le \mP_{i,k|k-1}. 
	\end{align*}
	Following the  map $\pi_k:\mP_{i,k|k-1} \mapsto \mP_{i,k|k}$ given by \eqref {eq:umv:cov} is operator monotone, $\mP_{i,k|k-1}^{\update} \le \mP_{i,k|k-1}$ indicates
	\begin{align*}
	\mP_{i,k|k}^{\update} \le \mP_{i,k|k}, \ \bar {\mP}_{i,k|k}^{\update} \le \bar{\mP}_{i,k|k}.
	\end{align*}
  Furthermore, following the map $\mathcal{F}_k: (\bar{\mP}_{1, k,k}, \dots, \bar{\mP}_{\nS, k,k}) \mapsto \mP_{k|k}$ given by \eqref{eq:CI:cov} is operator monotone, $\bar{\mP}_{i,k|k}^{\update} \le \bar{\mP}_{i,k|k}$ indicates
	$$ \mP_{k|k}^{\update} \le \mP_{k|k}. $$
	Then, Algorithm~\ref{alg:PP-ACF} remains $\mP_{i,k|k}$, while Algorithm~\ref{alg:PP-ACF:feedback} updates $\mP_{i,k|k}^{\update}$ via Step 4 and Step 5 using the local update strategy, which ensures
	\begin{align*}
	\mP_{i,k|k}^{\pre} \le \mP_{i,k|k}^{\update} \le \mP_{i,k|k}.
	\end{align*}
	This completes the proof.
$\hfill\blacksquare$

\begin{remark}
Propositions~\ref{prop:DP} and~\ref{prop:accuracy_enhancement} indicate that Algorithm \ref{alg:PP-ACF:feedback} enhances the fusion estimation accuracy of Algorithm \ref{alg:PP-ACF} while ensuring the same ($\epsilon,\delta$)-differential privacy. However, more computational cost is needed in Algorithm \ref{alg:PP-ACF:feedback}. 
Specifically, the computational complexity of Algorithm \ref{alg:PP-ACF:feedback} is $\mathcal{O}(\nS \dx^3)$ more than that of Algorithm \ref{alg:PP-ACF}. 
Overall,  Algorithm \ref{alg:PP-ACF} prefers to a scenario with low-complexity requirements, while Algorithm \ref{alg:PP-ACF:feedback} prefers to a  scenario with high-accuracy requirements.
\end{remark}

\begin{remark}
	Compared to encryption-based methods (see, e.g., \cite{Xu-2023-Distributed,Yan-2023-Distributed,Yan-2023-Guaranteeing}), the proposed algorithms are more efficient. Specifically, the computational cost is lower because it primarily involves solving an SDP, which can be performed efficiently using interior-point methods (see, e.g., \cite{wright1997primal}). Moreover, the communication overhead is lower because only noisy estimates and their covariance matrices should be transmitted. Beyond efficiency, a further distinction lies in the privacy guarantee: while encryption provides a binary guarantee reliant on computational assumptions, our differential privacy framework ensures a probabilistic guarantee.
\end{remark}

\begin{remark}
In implementing the proposed algorithms, two practical issues are addressed as follows: i) numerical stability,  which is ensured  via square-root filtering techniques (see, e.g., \cite{simon2006optimal}); and ii) potential model uncertainty, which is ensured via robust filtering techniques (see, e.g., \cite{simon2006optimal}). 
\end{remark}


\section{Example}\label{sec:example}

Consider the dynamic model in \cite{Bar-Shalom-Li-KirubarajanEstimation2001} as follows:
\begin{align*}
	\vx_{k + 1} = \mA_k \vx_k + \mB_k \vd_k + \vw_k, 
\end{align*}
where 
\begin{align*}
	\mA_k = \begin{bmatrix*}
	 1 & \Delta t & 0 & 0 \\
	 0 & 1        & 0 & 0 \\
	 0 & 0        & 1 & \Delta t \\
	 0 & 0        & 0 & 1
	\end{bmatrix*},
   \mB_k = \begin{bmatrix}
   	1 & 0 \\
   	0 & 0 \\
   	0 & 1 \\
   	0 & 0 \\
   \end{bmatrix}, 
   \vd_k = \begin{bmatrix}
   	5 \cos k  \\ 
   	5 \cos k 
   \end{bmatrix},
\end{align*}
and $\vw_k \sim \N(0, \mQ_k)$ with $\mQ_k = \diag([1,0.1,1,0.1]^{\Trans})$. The initial state is generated from $\N(\hat \vx_0, \mP_0)$ with $\hat \vx_0 = [0,5,0,5]^{\Trans}$ and $\mP_0 = \diag([10,10,10,10]^{\Trans})$. 
The distributed multi-sensor network system consists of two local sensors with their measurement model being
\begin{align*}
	\vy_{i,k} = \mC_{i,k} \vx_k + \vv_{i,k}, \ i = 1, 2,
\end{align*}
where 
\begin{align*}
\mC_{1,k} = \begin{bmatrix}
		1 & 0 & 0 & 0 \\
		0 & 0 & 1 & 0
	\end{bmatrix},\ 
    \mC_{2,k} = \mI_4,
\end{align*}
and $\vv_{i,k} \sim \N(0, \mR_i)$ with $\mR_1 = 0.1 \mI_2$ and $\mR_2 = 20 \mI_4$. 

\begin{figure}[htbp]
	\centering
	\includegraphics[width=0.465\textwidth]{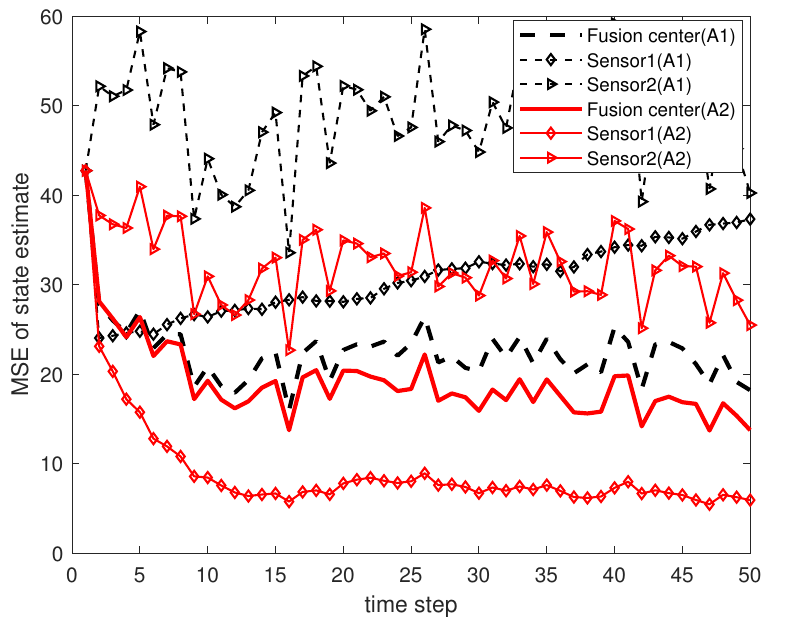}
	\caption{MSEs of local and fusion estimates for Algorithm \ref{alg:PP-ACF} and Algorithm \ref{alg:PP-ACF:feedback}.}
	\label{fig:1}
\end{figure}

Set $\epsilon_0 = 0.1$, $\epsilon= \delta= 10^{-3}$ for differential privacy level (see Definition \ref{def:DP}). 
In the covariance intersection fusion \eqref{eq:CI:mean} and \eqref{eq:CI:cov},  uniform weights ($w_1 = w_2 = 0.5$) are adopted for simplicity.
To demonstrate the effectiveness of the proposed two algorithms, 
the MSEs of local and fusion estimates over $50$ time steps and $50$ Monte Carlo runs for Algorithm \ref{alg:PP-ACF} (legends labeled A1) and Algorithm \ref{alg:PP-ACF:feedback} (legends labeled A2) are depicted in Fig.~\ref{fig:1}. 
We have the following observations and explanations:

1) Under the same differential privacy level, the MSEs of local and fusion estimates in Algorithm \ref{alg:PP-ACF:feedback} are smaller than those in Algorithm \ref{alg:PP-ACF} consistently. Intuitively, for the same line shape, the red lines are all below the black lines. This demonstrates that Algorithm \ref{alg:PP-ACF:feedback} does enhance the local and fusion estimation accuracy of Algorithm \ref{alg:PP-ACF}.

2) For Algorithm \ref{alg:PP-ACF}, the MSEs of fusion estimates are smaller than those of local estimates consistently. Intuitively, the black solid line is below the other two black marked lines consistently. This is due to the effectiveness of the covariance intersection approach adopted at the fusion center. 

3) For Algorithm \ref{alg:PP-ACF:feedback}, however, the MSE of fusion estimate is smaller than that of Sensor 2 and larger than that of Sensor 1 at each time step. Intuitively, the red solid line is between the other two red marked lines. This is because the covariance intersection approach is inherently conservative to ensure consistency (Remark \ref{rem:CI consistency}). As a result, Sensor 2 may not trust the feedback from the fusion center, leading to a lower local estimation accuracy compared with the fusion center. 

Table \ref{tab:averaged MSEs} lists the averaged MSEs for different weight selections in covariance intersection. The results show that Algorithm \ref{alg:PP-ACF:feedback} outperforms Algorithm \ref{alg:PP-ACF} across all weighting schemes. Moreover, the MSEs of both algorithms decrease as the weight $w_1$ increases. This trend is consistent with the higher measurement accuracy of Node 1 compared to Node 2, as reflected by the smaller order of magnitude of its measurement noise covariance $\mR_1$.

\begin{table}[htbp]
	\renewcommand{\arraystretch}{1.5}
	\caption{Comparison of averaged MSEs under different weight selections for covariance intersection.}
	\setlength{\tabcolsep}{1.5mm}
	\label{tab:averaged MSEs}
	\centering
	\begin{tabular}{c|c|c|c}
		\hline
		\diagbox{Algorithm}{MSE}{($w_1,w_2$)} & \makecell{($0.4,0.6$)} & \makecell{($0.5,0.5$)} & \makecell{($0.6,0.4$)} \\
		\hline
		Algorithm \ref{alg:PP-ACF}  & $34.36$  & $26.12$ & $24.67$  \\
		\hline
		Algorithm \ref{alg:PP-ACF:feedback}   & $22.43$ & $16.86$ & $11.85$   \\
		\hline
	\end{tabular}
\end{table}

\begin{figure}[htbp]
	\centering
	\begin{subfigure}[b]{0.24\textwidth}
		\includegraphics[width=\textwidth]{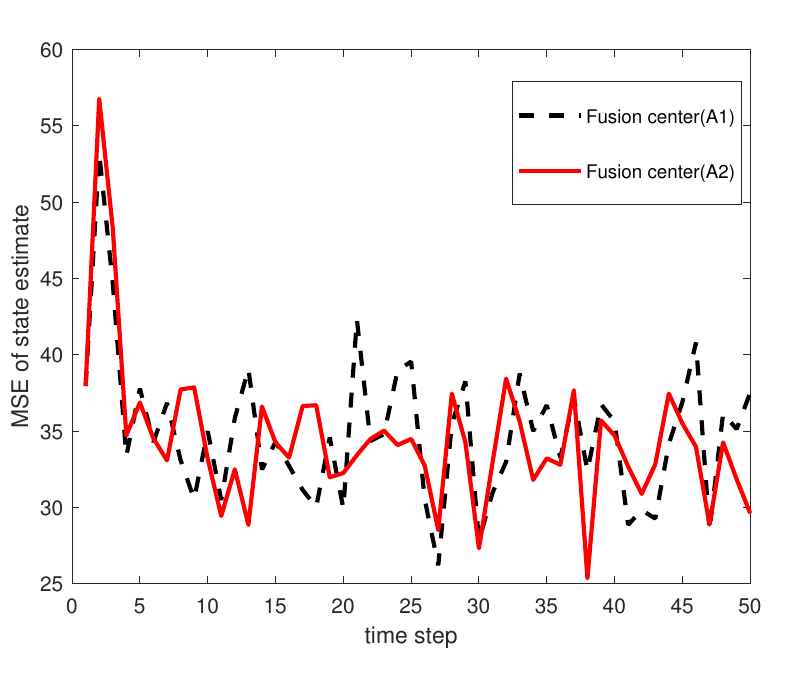}
		\caption{$\epsilon_0 = 0.5, \epsilon= \delta= 10^{-3}$.}
		\label{subfig:22}
	\end{subfigure}
    \begin{subfigure}[b]{0.24\textwidth}
	\includegraphics[width=\textwidth]{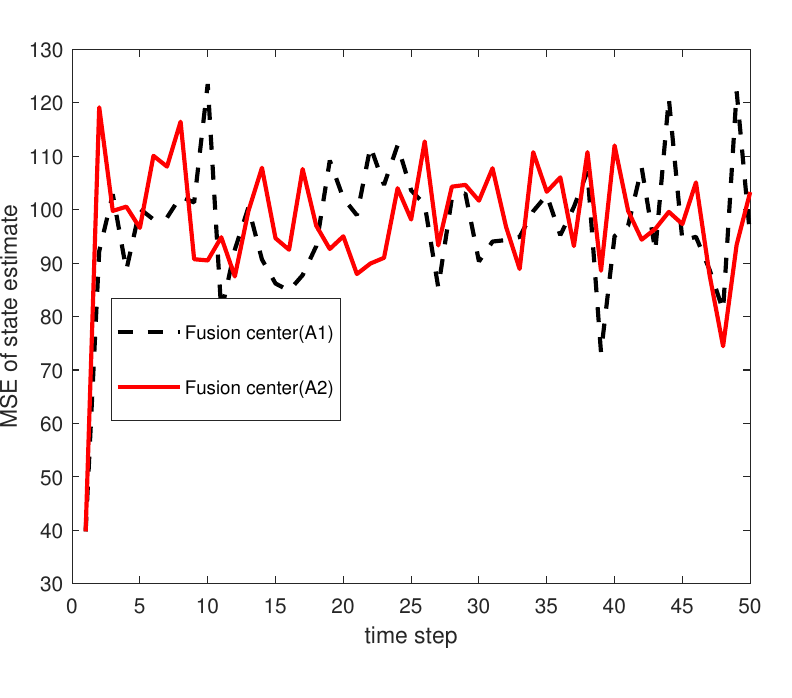}
	\caption{$\epsilon_0 = 1, \epsilon= \delta= 10^{-3}$.}
	\label{subfig:11}
    \end{subfigure}
	\caption{MSEs of fusion estimates for different $\epsilon_0$.}
	\label{fig:11}
\end{figure}

\begin{figure}[htbp]
	\centering
	\begin{subfigure}[b]{0.24\textwidth}
		\includegraphics[width=\textwidth]{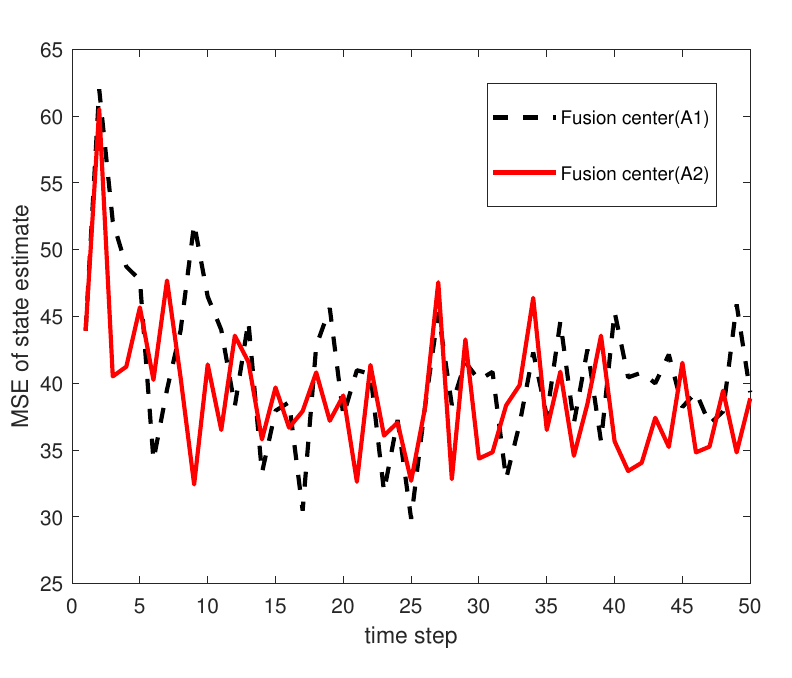}
		\caption{$\epsilon_0 = 0.1, \epsilon= \delta= 10^{-6}$.}
		\label{fig:4}
	\end{subfigure}
	\begin{subfigure}[b]{0.24\textwidth}
		\includegraphics[width=\textwidth]{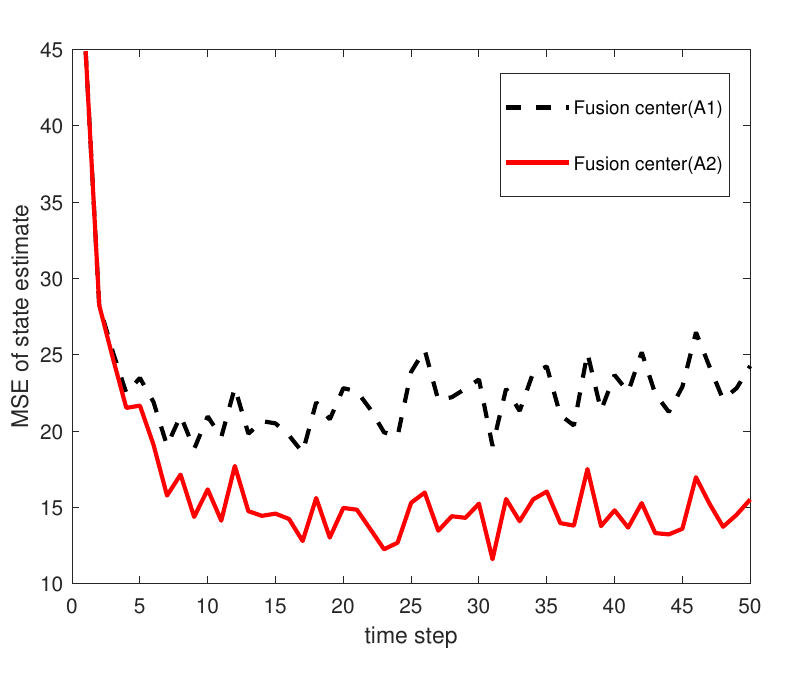}
		\caption{$\epsilon_0 = 0.1, \epsilon= \delta= 0.1$.}
		\label{fig:5}
	\end{subfigure}
	\caption{MSEs of fusion estimates for different $\epsilon$, $\delta$.}
	\label{fig:22}
\end{figure}

To show the trade-off between differential privacy level and fusion estimation accuracy, the MSEs of fusion estimates for different parameter settings are presented in Figs.~\ref{fig:11} and \ref{fig:22}. 

For different $\epsilon_0$, the MSEs of fusion estimates are compared in Fig.~\ref{fig:11}. Specifically, 
from Fig.~\ref{fig:11}(\subref{subfig:22}) to Fig.~\ref{fig:11}(\subref{subfig:11}), the $\epsilon_0$ becomes larger, and thus the differential privacy level gets higher. Nevertheless, the fusion estimation accuracy gets lower, as confirmed in Fig.~\ref{fig:11}(\subref{subfig:11}). 
For different $\epsilon$, $\delta$, the MSEs of fusion estimates are compared in Fig.~\ref{fig:22}. Specifically, 
from Fig.~\ref{fig:22}(\subref{fig:4}) to Fig.~\ref{fig:22}(\subref{fig:5}), the $\epsilon$ and $\delta$ becomes larger, and thus the differential privacy level gets lower. As expected, the fusion estimation accuracy in Fig.~\ref{fig:22}(\subref{fig:5}) is superior to that in Fig.~\ref{fig:22}(\subref{fig:4}). 
Overall, Figs.~\ref{fig:11} and \ref{fig:22} illustrate the trade-off between differential privacy level and fusion estimation accuracy.






\section{Conclusion}

In this paper, we have achieved  distributed fusion estimation while protecting the exogenous inputs against full eavesdropping. 
By solving a constrained minimization problem,
the ($\epsilon, \delta$)-differential privacy is ensured and the sum of MSEs of local estimates is minimized simultaneously.
To deal with the non-convexity of the minimization problem, we have relaxed it to the SDP problem and then solve it efficiently, making valuable sense in real-time state estimation.
Consequently, we have developed two differentially private distributed fusion estimation algorithms applicable to different scenarios: One prefers to low-complexity requirements, while the other prefers to high-accuracy requirements.  
Particularly, the second algorithm with the feedback mechanism enhances the fusion estimation accuracy of the first algorithm while ensuring the same ($\epsilon, \delta$)-differential privacy, despite sacrificing some acceptable computational complexity. 

Given that protecting a sequence of inputs is more general
than protecting only the latest one, a promising future
direction is to generalize our approach to protect the latest $k_1$
inputs ($k_1 \geq 2$).

\bibliographystyle{IEEEtran}
\bibliography{mybibfile}

\ifCLASSOPTIONcaptionsoff
  \newpage
\fi

\end{document}